\documentclass[onecollarge]{svjour2}
\usepackage{graphicx}
\usepackage{cite}
\usepackage{color}
\usepackage{amsfonts}

\begin{document}

\title{Nonequilibrium wetting}

\author{Andre Cardoso Barato}

\institute{Fakult{\"a}t f{\"u}r Physik und Astronomie, Universit{\"a}t W{\"u}rzburg, Am Hubland, 97074 W{\"u}rzburg, Germany\\ \email{barato@uni-wuerzburg.physik.de}}

\date{Received: date / Accepted: date}

\maketitle

\begin{abstract}
 
When a nonequilibrium growing interface in the presence of a wall is considered a nonequilibrium wetting transition may take place. This transition can be studied trough Langevin equations or discrete growth models. In the first case, the Kardar-Parisi-Zhang equation, which defines a very robust universality class for nonequilibrium moving interfaces, with a soft-wall potential is considered. While in the second, microscopic models, in the corresponding universality class, with evaporation and deposition of particles in the presence of hard-wall are studied. Equilibrium wetting is related to a particular case of the problem, it corresponds to the Edwards-Wilkinson equation with a potential in the continuum approach or to the fulfillment of detailed balance in the microscopic models. In this review we present the analytical and numerical methods used to investigate the problem and the very rich behavior that is observed with them.  
\keywords{Wetting transitions \and Surface growth models \and Kardar-Parisi-Zhang equation}
 
\end{abstract}

\def\xvec{{\vec x}}

\def\d{{\rm d}}
\def\te{{t_{\scriptscriptstyle \hspace{-0.3mm} e}}}
\def\tf{{t_{\scriptscriptstyle \hspace{-0.3mm} f}}}
\def\Ps{{P_{\scriptscriptstyle \hspace{-0.3mm} s}}}
\def\Pe{{P_{\scriptscriptstyle \hspace{-0.3mm} e}}}

\newpage

\tableofcontents

\newpage
\vspace{25mm}
\parskip 2mm 
\pagestyle{plain}

\section{Introduction}
\label{1}
\begin{figure}
\begin{center}	
\includegraphics[width=95mm]{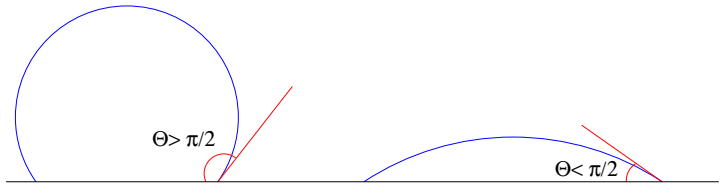}
\caption{Two drops on a substrate. The left one has a contact angle $\Theta>\pi/2$ and wets less the substrate, in comparison to the right one with $\Theta<\pi/2$.} 
\label{drops}
\end{center}
\end{figure}

Wetting \cite{landau} is well exemplified by considering a liquid droplet on a substrate. Depending on the physical properties of the system, determining the shape of the droplet, the substrate will be more or less wet. More specifically, the contact angle $\Theta$ (see Fig. \ref{drops}) is related to the surface tensions trough Young's equation,
\begin{equation}
\cos\Theta= (\sigma_{S,V}-\sigma_{S,L})/\sigma_{L,V},
\end{equation}       
where $\sigma_{S,V}$, $\sigma_{S,L}$ and  $\sigma_{L,V}$ are the surface tensions of the substrate-vapor, substrate-liquid  and  liquid-vapor surfaces, respectively. Total wetting happens if $\Theta=0$, and $0<\Theta<\pi$ corresponds to partial wetting.  

Cahn \cite{cahn77} observed that by approaching the critical temperature $T_c$, for $T<T_c$, the liquid-vapor surface tension $\sigma_{L,V}$ goes to zero faster than the difference $(\sigma_{S,V}-\sigma_{S,L})$, therefore, at a temperature $T_W<T_c$ a wetting transition should take place. In this wetting transition, $\Theta>0$ for $T<T_W$ and $\Theta=0$ for $T\ge T_W$. This observation introduced the notion that wetting could be viewed as a type of critical phenomena.   

More generally, a wetting transition occurs in a thermodynamic system constituted of a bulk phase $A$ and a substrate that attracts a second coexisting phase $B$. Depending on the control parameters, like e.g. temperature and chemical potential, the system will be in the moving phase or in the bound phase. At the wetting transition a macroscopic layer of the absorbed phase $B$ is formed and phases $A$ and $B$ coexist. In the bound phase only the phase $A$ is stable and the $AB$ interface stay pinned to the wall, whereas in the moving phase the $AB$ interface grows and only phase $B$ is stable.

Abraham \cite{abraham80} introduced a two-dimensional Ising model that could be solved exactly and displayed the same kind of transition predicted by Cahn \cite{cahn77}; it was also shown that in the solid on solid (SOS) limit the model still exhibited  a wetting transition. The advantage of taking this limit is that it simplifies the calculations and with it several exact results can be obtained using the transfer matrix method \cite{leeuwen81,burkhart81, abraham86}.

The SOS model, defined on a one-dimensional lattice with periodic boundary conditions and size $L$, has the following Hamiltonian,  
\begin{equation}
H= J\sum_{i=1}^L|h_{i+1}-h_i|+\sum_{i=1}^L V(h_i),
\label{1HSOS}
\end{equation} 
where $h_i\ge 0$ is a discrete random variable representing height, $J$ is a positive coupling, $V(h_i)$ is a potential accounting for the interaction between the wall ($h_i=0$) and the absorbed phase and no overhangs are allowed. The region $x>h_i$ corresponds to the  phase $A$ and the region $x<h_i$ to the coexisting phase $B$, therefore, the height gives the position of the $AB$ interface. By varying the temperature this model presents a wetting transition at a temperature $T_W$, at which the average height of the interface diverges.

On a coarse-grained level equilibrium wetting transitions can be studied using the Hamiltonian \cite{dietrich86}
\begin{equation}
H= \int d^{d}{\bf x}\{\frac{\sigma}{2}|\nabla h({\bf x})|^2+ V[h ({\bf x})]\}, 
\label{1H}
\end{equation}
where ${\bf x}$ is a continuous variable giving the position in the $d$-dimensional substrate. The height $h({\bf x})$ gives the position of the interface and the potential $V[h({\bf x})]$ accounts for the presence of the substrate and a possible interaction between it and the interface. In order to consider the dynamics of wetting, Lipowsky \cite{lipowsky85} introduced the following Langevin equation,
\begin{equation}
\frac{\partial h({\bf x},t)}{\partial t}= a+\sigma \nabla^2 h({\bf x},t)-\frac{\delta V[h({\bf x},t)]}{\delta h({\bf x},t)}+\zeta({\bf x},t),
\label{bEW}
\end{equation}
where the deterministic part of it originates from $-\delta H[h(x)]/\delta h(x)$, with $H$ given by (\ref{1H}), and $\zeta({\bf x},t)$ is a Gaussian noise with zero mean and variance given by
\begin{equation}
\langle\zeta({\bf x},t)\zeta({\bf x'},t') \rangle= D \delta({\bf x-x'})^{d-1}\delta(t-t').
\end{equation} 
Equation (\ref{bEW}) is  the Edwards-Wilkinson equation \cite{edwards82}, which is known to describes the motion of an equilibrium interface with velocity $a$ (see \cite{barabasi95, krug97}), with an extra term:  the potential $V[h(\bf x)]$.

The Kardar-Parisi-Zahng (KPZ) \cite{KPZ} equation differs from the EW equation by the presence of the nonlinear term $\lambda(\nabla h)^2$. The basic difference of an interface described by the KPZ equation (when compared to one described by the EW equation) is that the velocity of the interface depends on the local interface slope $\nabla h({\bf x})$. The non-linear term comes from an expansion of the velocity with respect to the local interface slope, and it can be shown that higher order terms in the expansion are irrelevant under a renormalization group transformation \cite{barabasi95, krug97}. That is why the KPZ equation defines a robust universality class for nonequilibrium moving interfaces. 

Since the paper of Cahn \cite{cahn77}, equilibrium wetting has been intensively studied, theoretically and experimentally (see  \cite{dietrich86, degennes85, gama85} for reviews). Differently, nonequilibrium wetting is in it's very beginning and experimental studies on it still lacks. One way to study it is to add the nonlinear term of the KPZ equation $\lambda(\nabla h)^2$ to equation (\ref{bEW}), such equation was considered for the first time by Tu et al. \cite{tu97}. Hinrichsen et al. \cite{hinrichsen97} and Mu\~noz and Hwa \cite{munoza98} introduced microscopic models in the KPZ universality class in the presence of a hard wall, showing that nonequilibrium wetting could also be studied with them. A considerable amount of work on nonequilibrium wetting has already emerged and our aim is to review it here.

In this article we present a detailed description of nonequilibrium wetting, reviewing the very rich phenomenology that is obtained from numerical and analytical methods used to study the Langevin equation and the microscopic models. Two reviews on the subject, more focused on the Langevin equation approach, are \cite{munoz04,santos04}, and there is some overlap between them and the present one. Nevertheless, in the account we make here many aspects of nonequilibrium wetting are presented in more detail, mainly when it comes to microscopic models. We also obtain a few new results that will be pointed out when they come.   

The review is organized in the following way. In the next section we define the Langevin equations and the microscopic models to be analyzed. Moreover, the quantities of interest and the exponents associated with them are defined. Sec. 3  contains exact calculations, that are possible for microscopic models when detailed balance is fulfilled. In Sec. 4   Monte Carlo simulations of microscopic models and numerical integration of the Langevin equations are presented. The content of Sec. 5 is mean field  approximations for microscopic models. The case of a substrate with more than one dimension is treated in Sec. 6, where mean field approximations for the continuum model, naive power-counting and some renormalization group arguments are used. We end with final remarks in Sec. 7.      

\section{Definition of the problem}
\label{2}
\subsection{Continuum model}
The Langevin equation we are going to consider, describing the motion of an interface in the presence of a wall, is the bounded KPZ (bKPZ) equation, given by \cite{tu97}
\begin{equation}
\frac{\partial h({\bf x},t)}{\partial t}= a-\frac{d}{dh}V(h)+\sigma\nabla^2 h({\bf x},t) +\lambda (\nabla h({\bf x},t))^2+\zeta({\bf x},t).
\label{bKPZ} 
\end{equation}   
This is equation (\ref{bEW}) added with the nonlinear term $\lambda(\nabla h({\bf x}, t))^2$. In contrast to the unbound KPZ interface, the sign of $\lambda$ is of great importance, because, as we will see, it leads to different universality classes. We call them bKPZ+ and bKPZ-- universality classes. 

The potential $V(h)$ has to account for the presence of a substrate and the interaction between the absorbed phase and the substrate. A natural way to consider the presence of a substrate is to forbid negative heights with a hard-wall potential, that is given by

\begin{equation}
V(h)= \left\{\begin{array}{cc}
0 & \textrm{if } h>=0\\
\infty & \textrm{if } h<0.
\end{array}\right.
\end{equation} 
The problem with this potential is that it is not suitable for calculations (analytical and numerical). A potential that overcomes this problem and is used in the study of equilibrium and nonequilibrium wetting with short range interactions is \cite{dietrich86} 
\begin{equation}
V(h)= \frac{b}{s}\exp(-sh)+ \frac{c}{2s}\exp(-2sh),
\label{2eqVh}
\end{equation}
where $s\ge 1$ controls the hardness of the wall, $b$ is a control parameter that in equilibrium and near the transition is proportional to $|T-T_W|$ and $c\ge 0$. If $b>0$ the wall is repulsive and the term $\exp(-2sh)$ is irrelevant, with $b<0$ and $c>0$ we have an attractive wall. The order parameter of the wetting transition is $n= e^{-h}$. In the bound phase the mean height is finite and the interface stays pinned to the substrate, therefore $n>0$. In the moving phase the mean height grows linearly with time and $n=0$. As we will see, in the microscopic models a hard-wall is considered, and the critical behavior of them is the same as the critical behavior of the bKPZ equation with the soft-wall potential (\ref{2eqVh}).   

It turns out that the bKPZ equation (\ref{bKPZ}), with $V(h)$ given by (\ref{2eqVh}), can be transformed into a Langevin equation with multiplicative noise. This is done with the Cole-Hopf transformation $n=e^{-h}$, which, in the case that the sign of the nonlinear term is negative, gives the following equation for the order parameter \cite{grinstein96},
\begin{equation}
\frac{\partial n({\bf x}, t)}{\partial t}= -\frac{d}{dn}V(n) + \sigma\nabla^2 n({\bf x},t) + n({\bf x},t)\zeta({\bf x},t),
\label{MN1}
\end{equation} 
where without loss of generality we set $\lambda= -\sigma$ and chose to interpret the Langevin equation in the Stratonovich sense (interpreting it in the Ito sense would just produce a shift in the factor multiplying the linear term \cite{gardiner}). The potential, as a function of $n$, is now given by
\begin{equation}
V(n)= \frac{a}{2}n^2+\frac{b}{2+s}n^{2+s}+\frac{c}{2+2s}n^{2+2s},
\label{2eqVn}
\end{equation}
where the linear term $-a n$ appearing in equation (\ref{MN1}) is now incorporated in the potential. This equation was introduced by Grinstein et al. \cite{grinstein96}, later it was pointed out that it was equivalent to the bKPZ equation \cite{tu97}. The universality class it defines is called multiplicative noise 1 (MN1),  obviously MN1 and bKPZ-- are the same universality classes. The MN1 equation is also related to the synchronization transition in coupled map lattices \cite{pikovsky94}, therefore, there is a relation between this transition and the wetting transition. We do not discuss this point here, references about it are \cite{ahlers02,munoz03,munoz04}.

By applying the same transformation to the bKPZ+ case (now with $\lambda=\sigma$), the resulting equation is 
\begin{equation}
\frac{\partial n({\bf x}, t)}{\partial t}= -\frac{d}{dn}V(n) +\sigma\nabla^2 n({\bf x},t)- 2\sigma\frac{(\nabla n({\bf x},t))^2}{n({\bf x},t)} + n({\bf x},t)\zeta({\bf x},t).
\label{MN2}
\end{equation} 
This equation is known as the multiplicative noise 2 (MN2) equation. The MN1 (\ref{MN1}) and MN2 (\ref{MN2}) equations will be very important in the course of this review, because they are more convenient in several situations. For example, integrating them numerically is simpler than integrating the bKPZ equation directly, which suffers by numerical instability problems \cite{newman96}.

With this map between the bKPZ equations and the MN equations it is easy to see what are the effects of considering an upper wall instead of a lower wall. An upper wall is implemented with the potential $V(h)= \frac{b}{s}\exp(sh)+ \frac{c}{2s}\exp(2sh)$ in equation (\ref{bKPZ}), in this case an negative (positive) $\lambda$ corresponds to the case of an lower wall with positive (negative) $\lambda$. This can be verified by performing the Cole-Hopf transformation, but now with $n=e^{h}$, in the bKPZ equation with an upper wall. The result is: with a positive (negative) $\lambda$ the obtained equation is MN1 (MN2). In this review we always consider a lower wall, also in the microscopic models.

Before going to the definition of the microscopic models we point out that, although the bKPZ equation is defined for $d$ spatial dimensions, in most of this review we will restrict to the case $d=1$, the exception being Sec. 6. For this reason, the microscopic models are defined, in the following, on a one-dimensional substrate.  
 
\subsection{Microscopic models}

\subsubsection{Restricted solid on solid model}

\begin{figure}
\begin{center}	
\includegraphics[width=105mm]{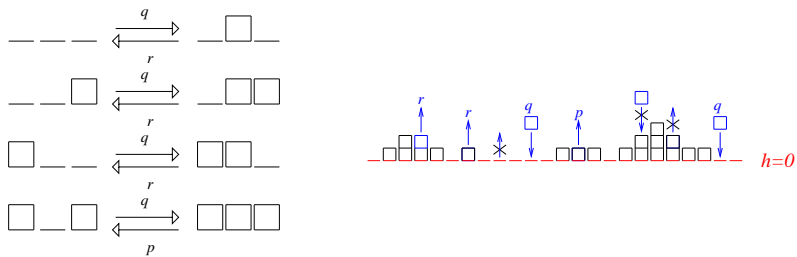}
\caption{Transition rates for RSOSW model (left) and example of an interface configuration with transitions that may take place.} 
\label{frates}
\end{center}
\end{figure}

The model for nonequilibrium wetting described in what follows was introduced by Hinrichsen et al. \cite{hinrichsen97}. It is a growth process taking place in a one-dimensional discrete lattice with periodic boundary conditions and size $L$. To each site $i$ a random variable $h_i$ is attached, it can take the values $h_i= 0,1,2,3...$ and is interpreted as the interface height. Nearest neighbors respect the restricted solid on solid (RSOS) constraint, i. e.,
\begin{equation}
|h_i-h_{i\pm 1}|\leq 1,
\end{equation}     
which introduces an effective surface tension. The interface evolves in time by random-sequential updates in the following way. A site $i$ of the lattice is randomly chosen and the processes that may occur are (see Fig. ~\ref{frates}):\\ 
\begin{itemize}
\item[(a)] deposition of a particle ($h_i\rightarrow h_i+1$) with rate $q$,\\
\item[(b)] evaporation of a particle ($h_i\rightarrow h_i-1$) at the edges of plateaus with rate $r$,\\
\item[(c)] evaporation of a particle ($h_i\rightarrow h_i-1$) from the middle of a plateau with rate $p$.\\
\end{itemize}
If the final configuration would violate the RSOS condition or would lead to a negative height, than it is not carried out. After $L$ attempts a Monte Carlo step is completed, and time is increased by one. The initial condition is a flat interface at height zero and we, without loosing generality, set $r=1$. A hard wall is present because evaporation events at the bottom layer ($h_i= 0$) are forbidden. Since this model respects the RSOS condition and there is a hard-wall, we name it RSOSW model.  

\begin{figure}
\begin{center}	
\includegraphics[width=85mm]{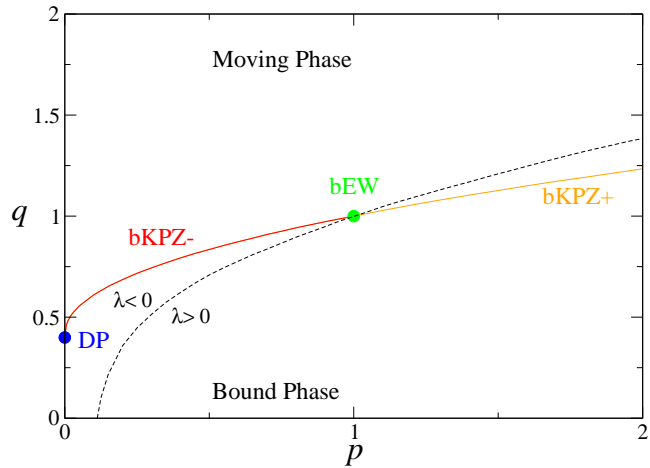}
\caption{Phase diagram of the RSOSW model. At $p=0$ the transition is in the direct percolation (DP) universality class; for $0<p<1$ it is in the bKPZ-- universality class; for $p=1$, where detailed balance is satisfied, it is in the bEW universality class; for $p>1$ in the bKPZ+ universality class.} 
\label{phasediagram}
\end{center}
\end{figure}

The wetting transition can be explained as follows. Consider a free interface (negative heights are allowed). It may propagate, depending on the deposition and evaporation rates, in the directions of increasing or decreasing height. In the phase where the velocity of the interface $v$ is positive (increasing height direction), the presence of the wall makes no difference: after some transient the interface will propagate with the same velocity as if the wall was not present. In contrast, when the rates are such that $v<0$, the wall changes the scenario completely because the interface stays bounded to the wall. Therefore, by forbidding negative heights, a wetting transition from a bound to a moving phase takes place, and the phase transition line $q_c(p)$ corresponds to a free interface with $v=0$. The phase diagram of the RSOSW model is displayed in Fig. \ref{phasediagram}.           

This model was generalized by Hinrichsen et al. \cite{hinrichsen00}, in order to include an attractive interaction between the substrate and the interface. This is done by considering a different deposition rate  $q_0<q$ at zero height. Because $q_0$ is smaller than $q$ the detachment of the interface from the substrate becomes harder and, therefore, this change in the dynamical rules simulates an attractive force between the substrate and the interface. As we will show, using microscopic models and the bKPZ equation, the presence of an attractive potential leads to new physics. 
  
The sign of the factor multiplying the nonlinear term of the the KPZ equation can be determined in a microscopic model. Considering a initially tilted interface, it is related to how the interface velocity varies with the tilt \cite{barabasi95}. In Fig. \ref{phasediagram} the doted-line where $\lambda=0$ is displayed, above (below) it $\lambda<0$ ($\lambda>0$). This shows that for $0<p<1$ the phase transition is in the bKPZ--, for $p=1$ in the bEW and for $p>1$ in the bKPZ+ universality class. $p=0$ is a particular case that will be addressed in Sec. 4.

\subsubsection{Single-step model}

Another microscopic realization of nonequilibrium wetting is the so-called  single-step model with a wall (SSW) introduced by Ginelli et al. \cite{ginelli03}. Here we use a more specific version of it, studied in \cite{ginelli04, kissinger05}. 

In the SSW model, the difference of height of two neighbors sites is restricted through the condition
\begin{equation}
|h_i-h_{i+1}|=1.
\end{equation}
As in the previous model, it is defined on a one-dimensional lattice of size $L$ and with periodic boundary conditions. It evolves random-sequentially by the following rules:  
\begin{itemize}
\item[(a)] deposition of a particle ($h_i\rightarrow h_i+2$) with probability $p$,\\
\item[(b)] evaporation of a particle ($h_i\rightarrow h_i-2$) with probability $1-p$,\\
\end{itemize}
at a chosen site $i$ occur if the configuration after the move is carried out does not violate the single-step constraint. The initial condition is $h_i= 1$ ($h_i=0$) if $i$ is odd (even). In the case of the SSW model $\lambda$ is proportional to $1/2-p$, therefore for $p=1/2$ we have the bEW case and for $p>1/2$ ($p<1/2$) the bKPZ-- (bKPZ+) case. The incorporation of a wall, in comparison to the previous case, is a bit more complicated, it is done in the following way. Different from the RSOSW model, it is not possible to vary the velocity of the interface keeping $\lambda$ fixed in the SSW model because there is just one control parameter, namely $p$. Fortunately, the velocity of the interface, in the long time limit, for the single-step model (free interface case) is known exactly; it is given by
\begin{equation}
v_L= (p-1/2)(1+1/L).
\label{2veloSS}
\end{equation}
Therefore, in order to study nonequilibrium wetting the system is tuned to criticality by considering a wall that moves with velocity $v_L$, given by (\ref{2veloSS}). This is done by forbidding evaporation events below  the substrate height $\bar{h}(t)$.

\begin{figure}
\begin{center}	
\includegraphics[width=155mm]{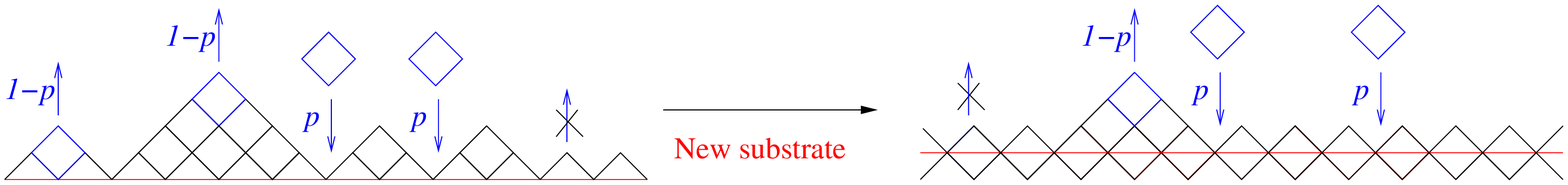}
\label{SSmodel}
\caption{Example of an interface configuration of the SSW model with some transitions that can occur. The horizontal line indicates the height of the moving wall $\bar{h}$. On the right is the interface after the substrate moves, $\bar{h}\to \bar{h}+1$.} 
\end{center}
\end{figure}

In this paper we attain ourselves to the cases $p=1/2$, $p=1$ and $p=0$ which are in the bEW, bKPZ-- and bKPZ+ universality classes, respectively. For $p=1/2$, since $v_L=0$, the height of the wall is $\bar{h}(t)=0$. At $p=1$, during a simulation, after  every $\Delta t=2(1-1/L)$ we increment the substrate height by one unity, $\bar{h}\to \bar{h}+1$. This means that all sites below the new $\bar{h}$ have their height increased and evaporation events at the new $\bar{h}$ are forbidden. For the bKPZ+ case ($p=0$) after every $\Delta t= 2(1-1/L)$ we have $\bar{h}\to \bar{h}-1$.

The SSW model is particular useful to obtain critical exponents numerically, since in this case one knows the critical point exactly. On the other hand the advantage of the RSOSW model is that it presents a richer behavior. The SSW model is also important in the study of mean field approximations (see Sec. \ref{5}).        

\subsection{Observables and related exponents}

Finally, we define the physical quantities, that we are going to consider in this review, and the exponents associated with them. They are defined in the context of the RSOSW model and will be generalized to the continuum model at the end of this section. The generalization to any other microscopic model is straightforward.  

\subsubsection{Scaling exponents}
A key quantity in what follows is the mean height of the interface, defined by
\begin{equation}
\langle h\rangle= \langle L^{-1}\sum_{i=1}^L h_i \rangle,
\end{equation}
where $\langle\rangle$ means the ensemble average. Another observable of great importance is the interface width, given by
\begin{equation}
w^2=  \langle L^{-1}\sum_{i=1}^L h_i^2 \rangle- \langle h\rangle^2.
\end{equation}

\begin{figure}
\begin{center}	
\includegraphics[width=95mm]{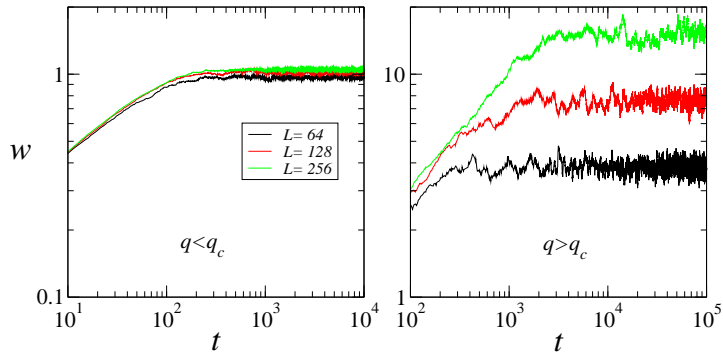}
\caption{Interface width $w$, for the RSOSW model, as a function of time in the bound phase (left) and in the moving phase (right) for $L= 64, 128, 256$. In the bound phase $q=0.9$ and in the moving phase $q= 3.0$, where $p=1.0$. One can see that the saturation value of $w$ tends to a constant in the bound phase (smooth interface) and grows with $L$ in the moving phase (rough interface).} 
\label{smoothrough}
\end{center}
\end{figure}

Considering an infinite system, if the interface width grows with time and does not reach a stationary value in the long time limit, the interface is called rough. Otherwise, if $w$ saturates after some transient it is smooth. In the one-dimensional case, the wetting transition is a roughening transition, the interface is smooth for $q<q_c$ and rough for $q\ge q_c$. Obviously, if the system is finite, even when the interface is rough, $w$ saturates. When one wants to verify if an interface is rough or smooth, by doing simulations in finite systems, one has to consider different sizes and see how the saturation value $w_s(L)$ varies with $L$. If $w_s(L)$ grows with $L$, than the interface is rough; if it tends to a saturation value independent of $L$, then the interface is smooth. This is shown in Fig. \ref{smoothrough}.
     
The KPZ universality class describes the self-affine properties of the roughening interface under scale transformations. The scaling exponents are defined by the relations \cite{barabasi95,krug97}
\begin{equation}
w(t)\sim t^\gamma,\qquad w_s\sim L^\alpha,\qquad t_s\sim L^z,
\label{scaling}
\end{equation}
where $t_s$ is the time at which the interface width saturates. $\gamma$ is known as the growth exponent, $\alpha$ is the roughness exponent and $z$ the dynamical exponent. They are not all independent, but related because of the Family-Vicsek
scaling relation \cite{family}
\begin{equation}
w(t,L)= L^\alpha f(tL^{-z}),
\end{equation} 
where $f(x)$ is a scaling function. From this last relation follows that $z= \alpha/\beta$. In one dimension the scaling exponents of the EW universality class are $\alpha= 1/2$, $\beta= 1/4$ and $z=2$; while for the KPZ universality class they are $\alpha= 1/2$, $\beta= 1/3$ and $z=3/2$ \cite{barabasi95,krug97}.   

We stress that the definition of the scaling exponents does not depend on the presence of the wall, they are related to the invariance of scale of $w$. However, the definition of critical exponents just make sense with a wall. As we will show, by introducing a wall, only one new independent critical exponent arises, the others can be determined by scaling relations and the values of the scaling exponents.     

\subsubsection{Critical exponents}

The order parameter of the wetting transition is the density of sites at zero height, i. e.,
\begin{equation}
\rho_0= \langle L^{-1}\sum_{i=1}^L\delta_{h_i,0}\rangle.
\end{equation}
In the bound phase $\rho_0>0$, while in the moving phase, where the interface detaches from the wall, $\rho_0=0$. Another order parameter for the transition is the velocity of the interface, which is zero at the bound phase and non-zero at the moving phase. 

\begin{figure}
\begin{center}	
\includegraphics[width=85mm]{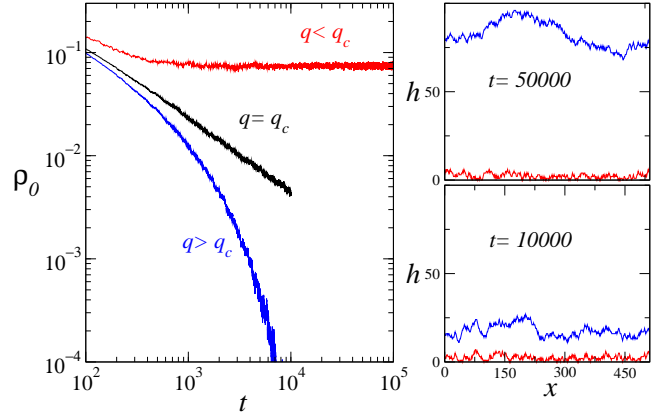}
\caption{On the left $\rho_0(t)$ below ($q=0.97$), at ($q=1.0$) and above ($q=1.01$) criticality, with  $L=512$ and $p= 1.0$. Note that at the critical point $\rho_0(t)$ decays algebraically, above it goes to zero faster exponentially and below it saturates. On the right, typical bounded and moving interfaces are showed for different times. For $q=0.97$ is stays bounded to the wall and for $q= 1.01$ it advances.} 
\label{grho0time}
\end{center}
\end{figure}

Fig. \ref{grho0time} shows the typical time evolution of $\rho_0$ above, below and at criticality. For $q<q_c$ the order parameter reaches a constant positive value in the stationary state. At the critical point, it goes to zero with a power-law behavior and for $q>q_c$ it vanishes exponentially. The exponent $\theta$ is defined at the critical line by the relation
\begin{equation}
\rho_0(t)\sim t^{-\theta}.
\label{theta}
\end{equation}
Near and below criticality we have
\begin{equation}
\rho_0^{s}\sim (q_c-q)^\beta,
\label{beta}
\end{equation}
where $\rho_0^s$ is the saturation value of the order parameter.
As discussed above the mean height is finite in the bound phase and diverges at criticality. The exponent associated with this divergence is defined by
\begin{equation}
\langle h \rangle\sim (q_c-q)^{-\zeta},
\label{zeta}
\end{equation}  
where the above relation is valid near and below criticality. In one dimension the interface width is also finite for $q<q_c$ and diverges at the critical point. Since it has the same dimension as $\langle h \rangle$ we expect it to diverge with the same exponent $\zeta$. In this sense, when dealing with a one-dimensional substrate, we consider the above definition also valid for $w$. When analytical calculations are possible, we calculate both quantities and show that, indeed,  they have the same critical behavior. 

At the wetting transition, like the interface width, the mean height follows the relation $\langle h\rangle\sim t^\gamma$. In the moving phase $\langle h\rangle$ grows linearly with time. Therefore, the interface velocity goes to zero as the critical point is approached from above. The critical exponent $\beta_v$ is defined by
\begin{equation}
v\sim (q-q_c)^{\beta_v},
\end{equation}     
where the above relation is valid above and near criticality. 

The spatial correlation length $\xi_\perp$ diverges near the critical point as,     
\begin{equation}
\xi_\perp\sim (q_c-q)^{-\nu_\perp}.
\label{nuperp}
\end{equation}
The same happens with the temporal correlation length $\xi_\parallel$, 
\begin{equation}
\xi_\parallel\sim (q_c-q)^{-\nu_\parallel}.
\label{nuparallel}
\end{equation} 
The saturation of the interface width $w$ in a finite system, for a rough interface, happens when $\xi_\perp$ becomes of the same order of the system size $L$, at the time $t_s$ proportional to $\xi_\parallel$. All this lead to 
\begin{equation}
\xi_\parallel\sim \xi_\perp^z,
\end{equation}
which gives the following scaling relation, 
\begin{equation}
z=\nu_\parallel/\nu_\perp.
\label{znuparallelnuperp}
\end{equation}
Another scaling relation is
\begin{equation}
\theta= \beta/\nu_\parallel,
\label{thetabetanuparallel}
\end{equation}
it comes from (\ref{theta}), (\ref{beta}) and (\ref{nuparallel}). Since the spatial correlation length is of the order of the system size when $w$ saturates, $w_s\sim L^\alpha$ can be written as $w_s\sim \xi_\perp^\alpha$. Hence, with (\ref{zeta}) and (\ref{nuperp}), we have 
\begin{equation}
\zeta= \nu_\perp\alpha. 
\label{scalingzetanuperp}
\end{equation}

We are considering the case $q_0=q$ where there is no attraction between the wall and the absorbed particles.
For the bKPZ equation, where the order parameter is $n= e^{-h}$, we have a non-attractive wall if $b$ assumes some positive fixed value (in this case the term $\exp(-2sh)$ is irrelevant and one can set $c=0$). The control parameter $a$ is  analogous to $q$ in the RSOSW model and, therefore, relations (\ref{theta}) and (\ref{beta}) become: $n(t)\sim t^{-\theta}$ (valid at $a=a_c$) and $n_s\sim (a_c-a)^\beta$, where $n_s$ is the saturation value.   

\subsubsection{Attractive substrate}

\begin{figure}
\begin{center}	
\includegraphics[width=125mm]{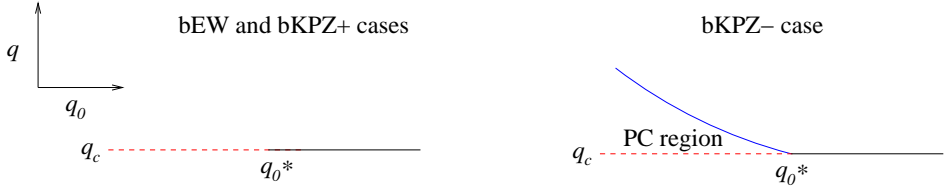}
\caption{Representation of the situations generated by the an additional attractive force between substrate and particles, for a fixed value of $s$. For the bKPZ+ and the bEW cases the transition goes from second order (full line) to first order (doted line) depending on the value of $q_0$. The critical point $q_c$ is not changed. For the bKPZ-- case there a is phase coexistence (PC) region. Considering the bKPZ equation, the situation is the same with $a$ playing the role of $q$ and $b$ the role of $q_0$.} 
\label{firstorderdraw}
\end{center}
\end{figure}

If there is an attractive force between the substrate and the particles, different physical properties are observed. Within the RSOSW model the situation is as follows. For the bEW and the bKPZ+ universality classes ($p\ge 1$), as the deposition rate at height zero $q_0$ decreases (more attraction), there is a threshold $q_0^*$, which depends on $p$, such that the transition becomes first order for $q_0<q_0^*$. In both cases the critical point $q_c$ remains unaltered, see Fig. \ref{firstorderdraw}.

In the bKPZ-- class ($0<p<1$) below $q_0^*$, a new critical value $q_c^{(2)}$ arises, this is shown in Fig. \ref{firstorderdraw}. For $q_0< q_0^*$ there is a phase coexistence region, in the sense that, depending on the initial conditions, the interface will be a moving or a bound one. For example, if a flat interface at a height far enough from the substrate is taken as initial condition, the interface will grow and not stay pinned to the substrate. Inside the phase coexistence region, the bound phase is the stable one in the thermodynamic limit, i. e., the average time for the interface to detach from the wall grows exponentially with the system size \cite{hinrichsen00,santos02,santos03}. The critical behavior of this new transition, taking place at value $q_c^{(2)}$, will be addressed in Sec. 4.   
   
Considering the bKPZ equation (\ref{bKPZ}) instead of the RSOSW model, the situation is the same as the one depicted in Fig. \ref{firstorderdraw},  with the parameter $a$ playing the role of $q$, the parameter $b$ the role of $q_0$ and $c$ with a fixed positive value. Note however that $a$ and $q$ or $b$ and $q_0$ do not have precisely the same meaning and their exact relationship is not known.  

In equilibrium wetting, at $a=a_c$, the transition that occurs when approaching $b_W$ (analogous to $q_0^*$), for $b< b_W$, is known as critical wetting. Whereas the transition taking place, for $b>b_W$, by approaching $a_c$, with $a< a_c$, is known as complete wetting \cite{dietrich86}. Therefore, critical (complete) wetting corresponds to $q=q_c$ ($q_0>q_0^*$) and $q_0$ ($q$) approaching $q_0^*$ ($q_c$) with $q_0<q_0^*$ ($q<q_c$) in the RSOSW model. As pointed out in \cite{santos03}, the transition taking place for $q_0<q_0^*$ by varying $q$  at the new critical point $q_c^{(2)}>q_c$ for the bKPZ-- class is not a wetting transition but rather a depinning transition because there is no phase coexistence at criticality  (phase coexistence corresponds to $q=q_c$, where the interface velocity is zero).  

At the tricritical point, $q_0= q_0^*$, new critical behavior is observed. We define the exponents associated to it using the superscript $t$. For example, at $q_0=q_0^*$, we have   
\begin{equation}
\rho_0^s \sim (q-q_c)^{\beta^t}.
\end{equation} 
We also define the following exponents associated to critical wetting, 
\begin{equation}
\rho_0^s\sim (q_0^*-q_0)^{\beta^{(2)}},
\end{equation}
\begin{equation}
w\sim (q_0^*-q_0)^{\zeta^{(2)}},
\end{equation}
where the above relations are valid at $q=q_c$.

\section{Exact results}
\label{3}
The solution of the RSOSW model, when it is on the bEW universality class, is presented below. For $p=1$ detailed balance holds and the model can be solved exactly for ($q\le 1$) \cite{hinrichsen97}. First we solve the model for $q_0=q$, then we consider the more general case $q_0\le q$ \cite{hinrichsen00}. Concerning exact solutions obtained with the height probability distribution in the stationary state in the bound phase, our presentation follows \cite{hinrichsen03a}. Also, by using a method introduced in \cite{neergaard97}, we calculate the velocity of a free interface for the RSOS model, leading to an exact calculation of the exponent $\beta_v$, which, to our knowledge, is not calculated elsewhere.

\subsection{Transfer matrix formalism}
\begin{figure}
\begin{center}	
\includegraphics[width=105mm]{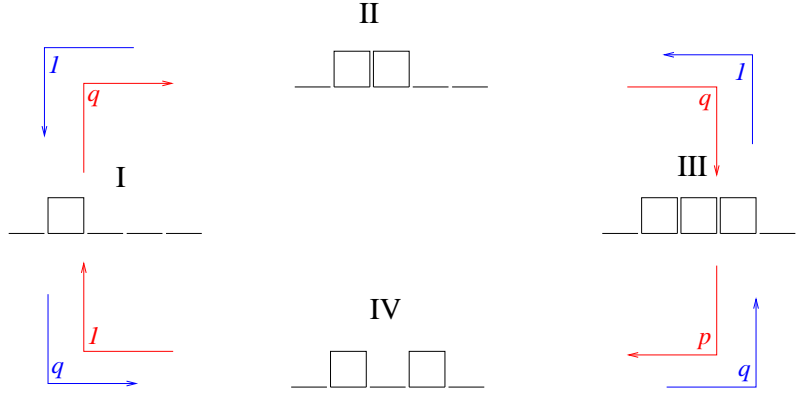}
\caption{Four different interface configurations forming a closed cycle and the respective transition rates.} 
\label{fdetailed}
\end{center}
\end{figure}

In a general dynamical system, detailed balance is fulfilled if, for every pair of microscopic $\sigma$ and $\sigma'$, the probability currents cancel each other, i. e.
\begin{equation}
P_\sigma w_{\sigma\to\sigma'}= P_{\sigma'}w_{\sigma'\to\sigma},
\label{eqDB}
\end{equation}
where $P_\sigma$ is the probability of being in the state $\sigma$ in the stationary state and $w_{\sigma\to\sigma'}$ is  the transition rate from $\sigma$ to $\sigma'$. From Fig. \ref{fdetailed} we see that, for the RSOSW model, it is satisfied only if
\begin{equation}
P_I= q^{-1}P_{II}= q^{-2}P_{III}= q^{-1}p^{-1}P_{IV}= p^{-1}P_{I},
\end{equation} 
which implies in $p=1$.

The detailed balance condition (\ref{eqDB}), with $p=1$, gives the following equation,
\begin{equation}
P(\{h_i\})= qP(\{h_i-1\}),
\label{eqDB2}
\end{equation}
where the configurations $\{h_i\}= h_1,...,h_i,..,h_L$ and $\{h_i-1\}= h_1,...,h_i-1,..,h_L$ are such that the RSOS constraint is obeyed. A simple ansatz arising from this condition is
\begin{equation}
P(\{h_i\})= Z_L^{-1}\prod_{i=1}^Lq^{h_i},
\label{eq3prob}
\end{equation}
where $Z_L$ is the partition function. It is given by
\begin{equation}
Z_L= \sum_{\{h\}}\prod_{i=1}^Lq^{h_i}, 
\end{equation}  
where the sum is over all configurations satisfying the RSOS constraint. We can write the probability distribution 
(\ref{eq3prob}) in the form
\begin{equation}
P(h_1,...,h_L)= Z_L^{-1}\prod_{i=1}^LT_{h_i,h_{i+1}},
\label{eq3probtranfer}
\end{equation}  
where $T_{h_i,h_{i+1}}$ is the transfer matrix and we are using periodic boundary conditions ($h_{L+1}= h_1$). The elements of the transfer matrix are given by 
\begin{equation}
T_{k,k'}= q^{(k+k')/2}(\delta_{k,k'}+\delta_{k,k'+1}+\delta_{k,k'-1}),
\label{eq3TM}
\end{equation}    
where $k\ge 0$ and $k'\ge 0$. Note that the transfer matrix is tridiagonal and this is a direct consequence of the RSOS constraint.   

Within the transfer matrix formalism, the partition function reads
\begin{equation}
Z_L= \sum_{h_1}...\sum_{h_L}T_{h_1,h_2}T_{h_2,h_3}...T_{h_{L-1},h_L}T_{h_L,h_1}= \sum_{h_1}T^L_{h_1,h_1}= \textrm{Tr}(T^L), 
\end{equation}
where the operator Tr gives the trace of the matrix. We are interested in calculating the density of sites at height $k$,
it can be written in the following form, 
\begin{equation}
\rho_k= Z_L^{-1}\sum_{\{h\}}\delta_{h_1,k}\prod_{i=1}^LT_{h_i,h_{i+1}}= Z_L^{-1}\sum_{h_2}...\sum_{h_L}T_{k,h_2}...T_{h_L,k}= Z_L^{-1} T^L_{k,k},  
\end{equation}
where the term $\delta_{h_1,k}$ imposes the constraint of summing only over configurations with $h_1=k$. In bra-ket notation the same quantity becomes
\begin{equation}   
\rho_k= Z_L^{-1}\langle k|T^L|k\rangle,
\end{equation}
where the vectors $|k\rangle$ form the canonical basis in height space.

In most of the following calculations the limit $L\to \infty$ will be taken: on it  
\begin{equation}
Z_L\approx \Lambda^L
\end{equation} 
and
\begin{equation}
\rho_k= \frac{|\langle k|\phi\rangle|^2}{|\langle\phi|\phi\rangle|},
\label{eqrho_h}
\end{equation} 
with $\Lambda$ being the maximum eigenvalue of the transfer matrix and $|\phi\rangle$ the corresponding eigenvector, i.e.
\begin{equation}
T|\phi\rangle= \Lambda|\phi\rangle.
\label{eq3eigen}
\end{equation}   

After setting up the transfer matrix formalism we proceed to calculate the critical exponents of the bEW universality class.

\subsection{Calculation of the critical exponents} 

Based on dimensional analysis, (\ref{beta}) and (\ref{nuperp}) we observe that the conditional correlation function,
\begin{equation}
c(l)= \frac{\langle\delta_{h_i,0}\delta_{h_i+l,0}\rangle}{\langle{\delta_{h_i,0}\rangle}},
\end{equation} 
is expected to follow the relation
\begin{equation}
c(l)\sim l^{-\beta/\nu_\perp},
\end{equation}
valid at the critical point. This quantity is equal to the sum of all possible paths, each of which multiplied by it's respective weight, connecting two points at height zero with a distance $l$ from each other. At the critical point all non-zero elements of the transfer matrix are equal to one, implying that all possible paths have the same weight. Therefore, for large enough $l$ and at criticality, the number of all possible paths connecting two points at height zero at a distance $l$ from each other is equivalent to the probability that a random walk starting at the origin will come back to the origin for the first time at time $l$. It is known that the probability distribution of the time that a random walker takes to return to the origin for the first time $\tau$ is given by $P(\tau)\sim \tau^{-3/2}$ \cite{redner}. From this follows that, at criticality,
\begin{equation}
c(l)\sim l^{-3/2}
\end{equation}  
and, therefore, $\beta/\nu_\perp= 3/2$. We note that this kind of random-walk argument was successfully used in equilibrium wetting \cite{fisher84}.

We now consider the problem in the thermodynamic limit, $L\to\infty$. From equations (\ref{eq3TM}) and (\ref{eq3eigen}) follows that  
\begin{equation}
q^k(q^{-1/2}\phi_{k-1}+\phi_{k}+q^{+1/2}\phi_{k+1})= \lambda\phi_{k},
\label{eq3eigen2}
\end{equation}
where $\phi_k$ is a component of the vector $|\phi\rangle$. In order to solve this equation we take the continuum limit, which should be valid when we are close enough to criticality. On this limit: $k\to h$, $\phi_k\to \phi(h)$ and equation (\ref{eq3eigen2}) becomes
\begin{equation}
\frac{d^2}{dh^2}\phi(h)+(3-\Lambda)\phi(h)-3\epsilon h\phi(h)=0, 
\end{equation}
where $\epsilon= 1-q$. The solution of it is
\begin{equation}
\phi(h)= Ai\bigg(\frac{3\epsilon h+\Lambda-3}{(3\epsilon)^{2/3}}\bigg),
\end{equation} 
where $Ai(x)$ is the Airy function. From the condition that $\phi(h)$ has to vanish for $h<0$ follows that $\Lambda=3$ and
\begin{equation}
\phi(h)= Ai(3^{1/3}\epsilon^{1/3}h).
\end{equation} 
With this explicit form of $\phi(h)$ the calculation of the mean height and the interface width is straightforward, they are given by
\begin{equation}
\langle h\rangle= A^{-1}\int_0^\infty\phi(h)^2h\sim \epsilon^{1/3}\qquad w= \sqrt{A^{-1}\int_0^\infty\phi(h)^2[h-\langle h\rangle]^2}\sim \epsilon^{1/3},
\end{equation}    
where $A= \int_0^\infty\phi(h)^2dh$.
Since $\phi(0)=0$, in order to calculate the density of sites at zero height in the continuum limit we have to calculate $\rho(h)$ at some small fixed height $h= \delta$. Considering that $\delta$ is much bigger than $\epsilon$ and is small enough such that $\phi(\delta)\approx \delta\phi'(0)$ is a good approximation, we have 
\begin{equation}
\rho(0)\sim A^{-1}(\phi'(0)^2),
\end{equation}
which gives
\begin{equation}
\rho(0)\sim \epsilon.
\end{equation}

These exact results are all for the stationary state, the present method does not allow us to calculate time-dependent quantities. In order to obtain them we resort to Monte Carlo simulations \cite{hinrichsen03a} or numerical integration of a set of equations obtained with the supposition that the time-dependent probability distribution is pair-factorized, which seems to be the case (see Sec. \ref{5}). Both methods give the following results,   
\begin{equation}
\langle h\rangle\sim t^{1/4},\qquad w\sim t^{1/4}
\end{equation}
and
\begin{equation}
\rho_0\sim t^{-3/4}.
\end{equation}

We have obtained the critical exponents of the bEW universality class for $d=1$, they are: $\beta=1$, $\nu_\perp=2/3$ and $\nu_\parallel= 4/3$. The above calculations are not possible for the bKPZ case ($p\neq 1$), where we have to use numerical simulations to obtain the critical exponents. Next we consider the new scenario generated by an attractive substrate for the bEW class.   

\subsection{The case $q_0\neq q$}  

With $q_0<q$, below a certain value of the deposition rate at height zero $q_0^*$ the phase transition becomes first-order. We now calculate the value $q_0^*$, show that the transition is first-order for $q_0<q_0^*$ and calculate the exponents associated to the tricritical point.

The only elements of the transfer matrix that are changed, when $q_0\neq q$, are $T_{0,0}= q/q_0$, $T_{0,1}=(q/q_0)^{1/2}$ and $T_{1,0}=(q/q_0)^{1/2}$. The new transfer matrix can be written in the form  
\begin{equation}
T_{k,k'}= q^{(k+k')/2}(q/q_0)^{(\delta_{k,0}+\delta_{k',0})/2}(\delta_{k,k'}+\delta_{k,k'+1}+\delta_{k,k'-1}),
\label{eq3TMq0}
\end{equation}    
where $k\ge 0$ and $k'\ge 0$. Using (\ref{eq3probtranfer}), one can verify that the probability distribution obtained with this transfer matrix satisfies detailed balance.

For $q>1$, the probability distribution is clearly not normalizable, independently of the value of $q_0$. This shows that the critical point is unchanged with the attractive force between the substrate and the absorbed particles. For $q=1$, we make the assumption that  
\begin{equation}
\phi_k= x^k\qquad \textrm{with }k\ge 1,
\label{eq3ansatzq0}
\end{equation}
where $x<1$. Using the transfer matrix (\ref{eq3TMq0}), the set of equations obtained by applying (\ref{eq3ansatzq0}) to equation (\ref{eq3eigen}) is
\begin{eqnarray}
q_0^{-1}\phi_0+q_0^{-1/2}x= \Lambda\phi_0,\nonumber\\
q_0^{-1/2}\phi_0+x+x^2= \Lambda x,\nonumber\\
x^{-1}+1+x=\Lambda.
\end{eqnarray}
It has the solution
\begin{equation}
\phi_0= q_0^{1/2},\qquad x= \frac{\sqrt{1+2q_0-3q_0^2}}{2(1-q_0)}-\frac{1}{2},\qquad \Lambda=\frac{x+1}{q_0}. 
\label{eqq0x}
\end{equation} 
Which, with equation (\ref{eqrho_h}) gives
\begin{equation}
\rho_0= \frac{1+q_0-6q_0^2+\sqrt{1+2q_0-3q_0}}{2+4q_0-6q_0}
\label{rho0q0}
\end{equation}
From this result we see that $\rho_0=0$ at $q_0^*=2/3$, $\rho_0>0$ for $q_0<q_0^*$ and for $q_0>q_0^*$ the assumption (\ref{eq3ansatzq0}) is not valid. Since $\rho_0$ is finite at the critical point $q=1$ for $q_0<q_0^*$, the transition is first-order.    

The exponents related to critical wetting can be obtained exactly as follows. From equation (\ref{rho0q0}), with $q_0$ below and near enough $q_0^*=2/3$, we have
\begin{equation}
\rho_0\sim (q_0^*-q_0)^1,
\end{equation} 
giving $\beta^{(2)}= 1$ for the bEW class. The density of particles at any height $h$ can also be calculated, from (\ref{eqrho_h}) and (\ref{eqq0x}), as follows
\begin{equation}
\rho_k= \frac{x^{2k}}{q_0+x^2/(1-x^2)}.
\end{equation}
The above equation with $\langle h\rangle= \sum k\rho_k$ and $w^2= \sum (k-\langle h\rangle)^2\rho_k$, gives
\begin{equation}
\langle h\rangle\sim (q_0^*-q_0)^{-1},\qquad w\sim(q_0^*-q_0)^{-1},
\end{equation}
therefore $\zeta^{(2)}=1$.

The critical exponents defined at the tricritical and it's vicinity, with $q<q_c$ and $q_0= q_0^*$, were obtained numerically in \cite{hinrichsen03a}. The results are in agreement with:  
\begin{equation}
\rho_0\sim (q_c-q)^{1/3},
\end{equation}
\begin{equation}
\langle h\rangle\sim (q_c-q)^{-1/3}\qquad w\sim (q_c-q)^{-1/3},
\end{equation}
\begin{equation}
\rho_0\sim L^{-1/2},
\end{equation}
where the first two relations are valid near and below criticality and the third at the critical point. The off-critical results can be confirmed by numerical diagonalization of the transfer matrix, whereas the finite-size result can be reproduced by evaluating of the product of $L$ transfer matrices. Time-dependent results, as in the case of complete wetting, are obtained by MC simulations \cite{hinrichsen03a} or numerical integration of the set of equations, obtained with the supposition that the time-dependent probability distribution factorizes (see Sec. 5), they are:
\begin{equation}
\langle h\rangle \sim t^{1/4},
\end{equation}
\begin{equation}
\rho_0\sim t^{-1/4}.
\end{equation}
These results give the exponents $\beta^t= 1/3$, $\nu_\perp^t= 2/3$ and $\nu_\parallel^t= 4/3$, different from the bEW critical exponents. Nevertheless, the exponents $z=2$ and $\gamma= 1/4$ are still the same.

\subsection{Exact calculation of the velocity of a free interface}

A method to obtain exact results, in the long time limit, for the RSOS model without the wall was developed by Neergaard and den Nijs \cite{neergaard97}. Here we use this method to calculate the velocity of a free interface and, consequently, obtain the exponent $\beta_v$. The calculations presented below are summarized in the appendix of \cite{alon98} for a slightly different model.  

The variables $\sigma_i= h_{i+1}-h_i$ can take only the values $-1,0,1$ because of the RSOS condition. With them the free interface problem can be mapped onto a problem of particles jumping in a lattice with the following rules:
\begin{eqnarray}
00\rightarrow+-\qquad\textrm{with rate }q\nonumber\\
0+\rightarrow+0\qquad\textrm{with rate }q\nonumber\\
-0\rightarrow0-\qquad\textrm{with rate }q\nonumber\\
-+\rightarrow00\qquad\textrm{with rate }q\nonumber\\
00\rightarrow-+\qquad\textrm{with rate }p\nonumber\\
0-\rightarrow-0\qquad\textrm{with rate }1\nonumber\\
+0\rightarrow0+\qquad\textrm{with rate }1\nonumber\\
+-\rightarrow00\qquad\textrm{with rate }1.
\label{3eqrules}
\end{eqnarray}
Since the initial condition is a flat interface and all possible transitions conserve the total charge, the number of positive charges ($\sigma_i= 1$) in the system is equal to the number of negative charges ($\sigma_i= -1$). In \cite{neergaard97} it was shown (in a more general context) that in the region of the phase diagram where
\begin{equation}
p= q\frac{2-q}{2q-1}\qquad\textrm{and}\qquad q>1
\end{equation}      
the  ansatz    
\begin{equation}
P(N)= Z_L^{-1} \bigg(\frac{q}{2q-1}\bigg)^{-N}, 
\label{P(N)}
\end{equation}
where $P(N)$ is the probability of having a configuration with $N$ positive charges in the limit $t\to \infty$ and $Z_L$ is a normalization constant, is in agreement with the dynamical rules (\ref{3eqrules}). The normalization condition gives
\begin{equation}
Z_L= \sum_{N=0}^{L/2}\frac{L!}{N!N!(L-2N)!}\bigg(\frac{q}{2q-1}\bigg)^{-N}.
\end{equation} 
In the thermodynamic limit, $L\to \infty$, the sum in the partition function $Z_L$ is dominated by the term $N= L/(2+1\sqrt{q/(2q-1)})$, giving the following density of positive (or negative) charges:
\begin{equation}
\rho= \frac{1}{2+\sqrt{(2q-1)/q}}+ o(1/L).
\label{eqrhorho}
\end{equation}   
The interface velocity is given by
\begin{equation}
v= (q-p)\langle00\rangle+ q\langle0+\rangle -\langle+0\rangle -\langle-0\rangle +q\langle-0\rangle
+q\langle-+\rangle-\langle+-\rangle.
\end{equation}   
Because of (\ref{P(N)}), $\langle00\rangle= (1-2\rho)^2+ o(1/L)$, $\langle+0\rangle= \langle0+\rangle= \langle-0\rangle=\langle0-\rangle= (1-2\rho)\rho+ o(1/L)$ and $\langle-+\rangle= \langle+-\rangle= \rho^2+ o(1/L)$. Hence,  
\begin{equation}
v= (q-1)2\rho+o(1/L),
\end{equation}   
where $\rho$ is given by (\ref{eqrhorho}).

We obtained the asymptotic velocity for the free interface case, which is equal to the asymptotic velocity of the RSOSW model in the moving phase. Even though the exponent $\beta_v$ is defined in the horizontal direction in the $q\times p$ plane we expect it to be the same in other directions, because numerical calculations of the exponent $\beta_v$ for different values of $p$ are all compatible with $\beta_v=1$ \cite{hinrichsen97}. Therefore, we have computed the exponent $\beta_v$ exactly.     

\section{Numerical simulations}
\label{4}
Here we discuss some technical aspects of Monte Carlo simulations of the RSOSW model and summarize numerical results for the bKPZ-- universality class obtained in \cite{barato08}. We then establish a scaling relation between the exponents, based on heuristic arguments and in agreement with the numerical results, which shows that with the introduction of the wall just one new independent exponent arises. We explain how to integrate the bKPZ equation numerically in an efficient way and demonstrate numerical results based on this method. The attractive wall case and extensions of the problem are also discussed.     

\subsection{The exponents of bKPZ-- universality class}   
In order to calculate the critical exponents numerically by simulating a lattice model one can use off-critical, finite-size and time-dependent simulations. With off-critical simulations it is possible to calculate the exponents $\beta$ and $\zeta$, while the exponents $\theta$ and $\nu_\perp$ can be obtained obtained with time-dependent and finite-size simulations, respectively.  

\begin{figure}
\centering
\includegraphics[width=85mm]{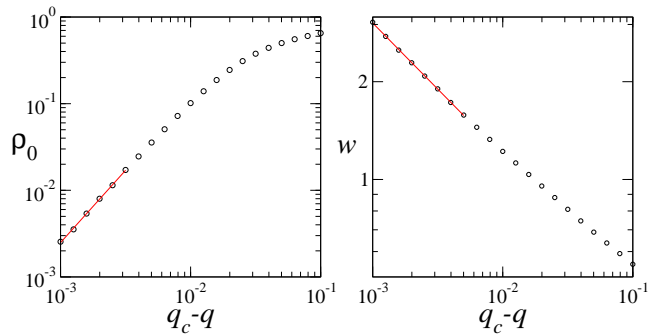}
\caption{Off-critical simulations. Density of sites with height zero $\rho_0$ (left) and the interface width $w$ (right)
as functions of the distance from the critical point $q_c= 0.4295(1)$, with $p=0.001$ and $L=4096$. Data taken from \cite{barato08}.} 
\label{goff}	
\end{figure}

A technical problem with the off-critical simulations is the EW-KPZ crossover, which manifests itself as follows. For a moderate simulation effort, it gives the impression that the critical exponents $\beta$ and $\zeta$ (measured for a fixed value of $p$) depended continuously on $p$, varying from the EW exponent $\beta=1$ and $\zeta=1/3$, when $p$ is near $1$, to larger values as $p$ gets smaller, where we are considering the bKPZ-- class ($0<p<1$). But this is not the case, when making measures near the equilibrium point one has to access regions closer to criticality in order to get appropriates value for $\beta$ and $\zeta$, they first look like bEW exponents and then they crossover slowly to the bKPZ-- values. Because of this crossover, simulations with small values of $p$ provide better results, that is why in \cite{barato08} the simulations were done at $p=0.001$. The off-critical simulations are presented in Fig. \ref{goff} and from it results 
\begin{equation}
\beta= 1.67(5)\,,\qquad \zeta= \nu_\perp\alpha= 0.41(5)\,.	
\end{equation}
The EW-KPZ crossover also takes place when considering the bKPZ+ case ($p>1$), with the difference that in this case, in order to get better numerical results, one has to use large values of $p$.

\begin{figure}
\begin{center}
\includegraphics[width=85mm]{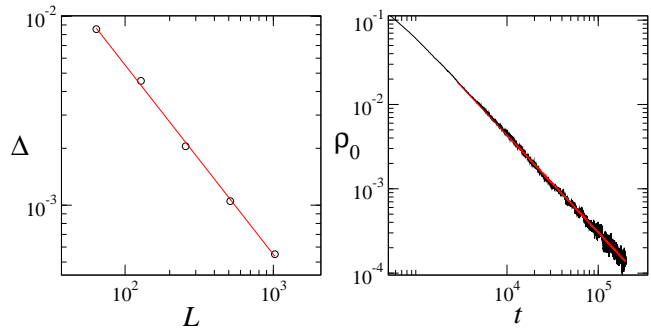}
\caption{Left: Difference $\Delta$ between the finite-size critical point $q_c(L)$ and the extrapolated critical point $q_c(\infty)= 0.4295(3)$ as a function of $L$ for $p=0.001$. Right: $\rho_0$ as a function of $t$ at the critical point $q_c= 0.4295(1)$, with $p=0.001$, $L=8192$ and $800$ realizations. Data taken from \cite{barato08}.} 
\label{gfin}	
\end{center}
\end{figure}

As for finite-size simulations, what is appropriate for the RSOSW model is to measure how the critical point varies with the system size, because, as observed in \cite{barato08}, this variation is very pronounced. From relation (\ref{nuperp})  it is expected that
\begin{equation}
q_c(\infty)-q_c(L)\sim L^{-1/\nu_\parallel},
\end{equation}  
where $q_c(L)$ is the critical point for a system of size $L$ and $q_c(\infty)$ is the extrapolated value. With the use of this relation, Fig. \ref{gfin} gives 
\begin{equation}
\nu_\perp= 1.00(3).
\end{equation}   
From the scaling relation (\ref{scalingzetanuperp}) and the above numerical result we get the exponent $\zeta= 0.5(1)$, in agreement with the value coming from off-critical simulations $0.41(5)$. 

With time-dependent simulations the exponent $\theta$ is obtained by plotting $n_0$ as a function of time at criticality. From Fig. \ref{gfin},      
\begin{equation}
\theta= \beta/\nu_\parallel= 1.15(3).
\end{equation}      
From the scaling relations (\ref{znuparallelnuperp}), (\ref{thetabetanuparallel}), the finite-size result $\nu_\perp= 1.00(3)$ and $z=3/2$ one can see that $\theta= 1.15(3)$ is in agreement with the off-critical result $\beta= 1.67(5)$.  

Results coming form finite-size and time-dependent simulations are certainly more reliable than when extracted form off-critical simulations. Nevertheless, off-critical simulations are important to confirm scaling relations. Next we propose a scaling relation and show that it is in agreement with the numerical results presented here.   

\subsection{Scaling picture and differences between the bKPZ universality classes}   

We now show, with an heuristic argument, that with the introduction of the wall just one new independent exponent arises. We suppose that the velocity of the interface in the growing phase follows $v\sim q-q_c$, which is in agreement with numerical results and an exact result presented in the last section. Since it is the time derivative of the mean height, which has the same dimension as the interface width $w$, and $\xi_\parallel$ has the dimension of time, we expect that $w\sim (q-q_c)\xi_\parallel$. With this, (\ref{zeta}), (\ref{nuperp}) and (\ref{znuparallelnuperp}) we have the scaling relation
\begin{equation}
\nu_\perp= \frac{1}{z-\alpha}.
\label{handwaving}
\end{equation}
In this argument we combined relations valid above criticality ($v\sim q-q_c$) with relations valid below. However, one can see in table \ref{table1}, where the exponents of the bKPZ and bEW universality classes are displayed, that the scaling relation (\ref{handwaving}) leads to exponents in agreement with the numerical (bKPZ classes) and exact (bEW class)  results obtained here. A possible reason for that is: without a wall for $q<q_c$ the interface would have a negative velocity, going linearly to zero as a function of the distance from the critical line. It seems that, with the presence of the wall, quantities with the dimension of the mean height divided by quantities with dimension of time still go to zero linearly as criticality is approached. The scaling relation (\ref{handwaving}), was obtained for the first time in \cite{tu97}, for all dimensions, using a different argument.  

With the scaling relation (\ref{handwaving}) and $z= \alpha/\gamma= \nu_\perp/\nu_\parallel$ we have that the critical exponents $\nu_\parallel$ and $\nu_\perp$ are determined by the scaling exponents. Therefore, there is only one independent (from the scaling exponents)  critical exponent left, which is $\beta$.  
 
\begin{table}[t]
\centering
\vspace{+0.1cm}
\begin{tabular}{|c|c|c|c|c|c|c|}
\hline
case 	      &$z$  &$\nu_\perp$&$\nu_\parallel$&$\zeta$&$\theta$   & $\beta$ \\
\hline
DP	      &$1.58$ &$1.10$& $1.73$& $0$  &$0.159$& $0.276$\\  
bKPZ--        &$3/2$&$1$ 	  &$3/2$  	  &$1/2$  &$1.184(10)$& $1.776(15)$\\
bEW 	      &$2$  &$2/3$      &$4/3$  	  &$1/3$  &$3/4$      & $1$\\
bKPZ+         &$3/2$&$1$        &$3/2$  	  &$1/2$  &$0.228(5)$ & $0.342(8)$\\
\hline
\end{tabular} 
\caption{List of the critical exponents. Most of DP exponents in the first line come from \cite{marrodickman}, the exceptions are $\alpha$ and $\zeta$ that come from \cite{hinrichsen03b}. The bEW exponents come from the exact results presented in Sec. 3. The bKPZ exponents come the numerical results for the exponent $\theta$ obtained in
\cite{kissinger05} and the scaling relations presented here. Table taken from \cite{barato08}} 
\label{table1}
\end{table}

All the above discussion should also be valid to the bKPZ+ universality class, with the difference that the new critical exponent $\beta$ (or $\theta= \beta/\nu_\parallel$) has a different value. The best way to obtain the new critical exponent numerically is to perform time-dependent simulations with the SSW model, since in this model the critical point is known exactly. This was done in \cite{kissinger05}, and the estimated exponents are summarized in table \ref{table1}.    

\begin{figure}
\begin{center}
\includegraphics[width=105mm]{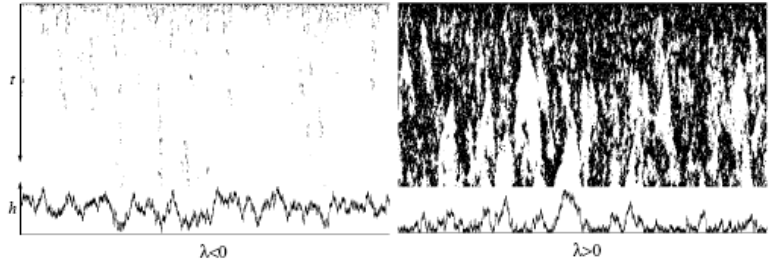}
\caption{Upper row: typical spatio-temporal configuration of contact points with the substrate. Lower row: final interface configuration at $t=5000$. The figures were obtained with numerical simulations of the SSW model at$ p=1$ ($\lambda<0$) and $p=0$ ($\lambda>0$). The number of contact points in the $\lambda<0$ case is clearly smaller. Figure taken from \cite{kissinger05}. 
} 
\label{diffig1}
\end{center}
\end{figure}

The exponent $\beta$ is bigger (smaller) than one for the bKPZ-- (bKPZ+) universality class, this tell us that a typical interface configuration of the bKPZ-- universality class, at criticality, is characterized by a smaller number of contact points with the substrate, in comparison to a bKPZ+ typical interface. In Fig. \ref{diffig1} we show typical interface configurations, obtained in \cite{kissinger05} with the SSW model at $p=0$ ($\lambda >0$) and $p=1$ ($\lambda <0$). For the bKPZ-- case, a smaller number of contact points and larger detached regions (distance between two contact points) are observed.  

Remarkably, the distance between the contact points differs from the correlation length. More specifically, since the average distance of contact points $l(t)$ is proportional to $\rho_0^{-1}(t)$ and $\nu_\perp=1$, we have that $l(t)> \xi_\perp(t)$ ($l(t)< \xi_\perp(t)$) for the bKPZ-- (bKPZ+) class. The interplay of these two length scales leads to other differences between the bKPZ+ and bKPZ-- universality classes, they are related to the distribution function of the distances between contact points and to the first depinning time probability distribution \cite{kissinger05}.

\subsection{Wetting as a Contact Process with an external field}

Let us now turn to the special case $p=0$. Here the RSOSW model exhibits a very different dynamics, which was introduced in \cite{alon96} and further analyzed in \cite{alon98, hinrichsen03b}. This is a very particular case of the model, because, with $p=0$, once a layer is completely filled an evaporation on it becomes impossible. Hence, since the initial condition is a flat interface at height zero, the presence of the wall makes no difference.

It was shown in \cite{alon96} that the phase transition at $p=0$ pertains to the direct percolation (DP) universality class \cite{hinrichsenreview,hinrichsenbook}, which is the most prominent universality class of nonequilibrium phase transitions into an absorbing state \cite{odor, odor2}. An explanation for this comes from the fact that the RSOSW model at $p=0$ can be related to the contact process (CP)\cite{marrodickman}. The CP is a well-known model in the DP universality class, it can viewed as a simple, and also simplistic, model for the propagation of a disease, where each site can be in two states: empty (healthy) or occupied by a particle (infected). Particles can create other particles in empty first neighbors sites (propagation of the disease) or die spontaneously (cure). If the system has no particles (no sick individuals) it is in the absorbing state and the dynamics ceases. An non-exact map between the CP and the RSOS at $p=0$ model was proposed in \cite{alon96} where sites with zero height are related to the infected individuals in the contact process. Hence, once the first layer is filled it is analogous to enter the absorbing state in CP, since no particles in this layer can be evaporated anymore and the height zero becomes inaccessible.            

In \cite{barato08} it was shown that the region in the phase diagram with $0<p<1$ can be interpreted as a DP process with an external field that destroys the transition. In the CP, an external field is introduced by allowing creation of particles at a certain rate. With it, there is no absorbing state anymore and the transition is lost. In the same way the non-zero $p$ destabilizes the absorbing state in the RSOSW model, since even after the first layer is completely filled an evaporation on it is still possible. With this interpretation the curvature of the phase transition line, Fig.\ref{phasediagram}, is predicted and a crossover exponent from the DP to the bKPZ-- universality class calculated (see \cite{barato08} for details). 

\subsection{Numerical integration of the bKPZ equation}
 
The MN equations can be integrated numerically very easily by using a method introduced in \cite{dornic05} to integrate Langevin equations with non-additive noise. The method consists of integrating the deterministic and stochastic parts of a discrete version of the Langevin equation separately in each time step. 

\begin{figure}
\begin{center}
\includegraphics[width=85mm]{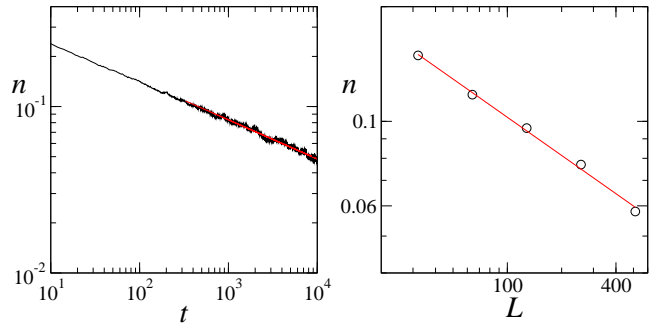}
\caption{Numerical integration of equation (\ref{nonorder}). Left: time-dependent simulations with the order parameter $n= \langle m\rangle$ as a function of time, at the critical point $a_c= 0.39025(25)$ for $L=2048$, giving $\theta= 0.235(10)$. Right: finite-size simulations with the saturation value of $n$ as a function of the system size $L$ at the critical point, giving $\beta/\nu_\perp= 0.33(2)$. The values of the other parameters are $b=1$, $c=0$, $\sigma= 0.1$, $D= 1$, $s= 4$ and $dt=0.1$.} 
\label{MN2integration}
\end{center}
\end{figure}

We now explain this method for the MN1 equation in the case of a one-dimensional substrate. To this end we have to consider the spatially discrete version of equation (\ref{MN1}), which reads
\begin{equation}
\frac{d}{dt}n_i= -a n_i -bn_i^{s+1} -cn_i^{2s+1}+ \sigma[n_{i+1}+n_{i-1}-2n_i]+n_i\zeta_i,  
\label{MN1d}
\end{equation}
where $n(x,t)\to n_i(t)$, $x\to i\delta x$ and we are using $\delta x= 1$. The algorithm evolves as follows. First a variable $n^*_i(t)$ is calculated from
\begin{equation}
n^*_i(t)= n_i(t)+ \{-bn_i(t)^{s+1}- cn_i(t)^{2s+1}+\sigma[n_{i+1}(t)+n_{i-1}(t)-2n_i(t)]\}dt,
\end{equation}
which corresponds to one step of the integration, using the Euler method, of the deterministic part (without the linear term) of equation (\ref{MN1d}). After that, $n_i(t+dt)$ is obtained with  
\begin{equation}
n_i(t+dt)= n^*_i(t)\exp(-adt+\sqrt{Dt}\eta_i),
\end{equation}
which is the solution of the one-variable Langevin equation constituted only of the stochastic part and the linear part of equation (\ref{MN1d}), where $\eta_i$ is a random number coming from a Gaussian distribution with zero mean and unitary variance. The reason to add the linear term in the second step of the update scheme is that the stochastic part can still  be solved exactly with it. Note that the term $-2\sigma n_i$, included in the first step, could, instead, be included in the second step. After all the $L$ variables are actualized according to the above scheme a step $dt$ is completed.      

The numerical integration of the MN2 equation is more complicated because of the term $(\nabla n)^2/n$ \cite{munoz03br}. In order to overcome this problem, Al Hammal et al. \cite{hammal05} considered a non-order parameter Langevin equation. Such an equation is obtained with the variable $m=1/n$ that transforms the MN2 equation (\ref{MN2}) into:
\begin{equation}
\frac{\partial}{\partial t}m= am+ bm^{1-s}+ cm^{1-2s}+ \sigma\nabla^2 m+ m\zeta.
\label{nonorder}
\end{equation}
We integrated this equation numerically using the algorithm explained above. We performed time-dependent and finite-size simulations, obtaining $\theta= 0.235(10)$ and $\beta/\nu_\perp= 0.33(2)$ (see Fig. \ref{MN2integration}). The results are in agreement with the results obtained in \cite{hammal05} and the results presented in table \ref{table1}. We note that  changing the value of $s$ does not change the critical behavior \cite{hammal05} (in Fig. \ref{MN2integration} we used $s=4$). 

\subsection{Numerical results for the attractive substrate case}

As discussed in Sec. 2, in the bKPZ-- case with an attractive force between the substrate and the interface there is a phase coexistence region if the attraction is strong enough and a depinning transition takes place at a new critical point (see Fig. \ref{firstorderdraw}). 

\begin{figure}
\begin{center}
\includegraphics[width=85mm]{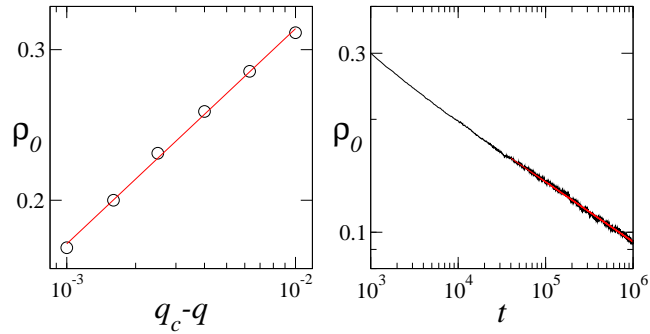}
\caption{Simulations of the RSOSW model with an attractive force between the substrate and the interface. Left: off-critical simulations giving $\beta=0.25(2)$. Right: time-dependent simulations giving $\theta= 0.159(10)$. Both exponents are in agreement with the DP exponents. The simulations were done at $p= 0.001$ and $q_0= 0.39$, where $q_c= 0.4295(1)$ and $q_c^{(2)}= 0.4377(1)$.} 
\label{gMNDP}
\end{center}
\end{figure}

This depinning transition was first observed to be first-order \cite{hinrichsen00, santos02}. Later Mu\~noz and Pastor-Satorras \cite{munoz03}, with numerical integration of the MN1 equation, obtained exponents of the DP universality class. In Fig. \ref{gMNDP} we present off-critical and time-dependent simulations for the RSOSW that agree with the findings from \cite{munoz03}. We used the values $p=0.001$, where $q_c= 0.4295(1)$, and $q_0= 0.39$ (which is smaller than $q_0^*$). We obtained  the new critical point at $q_c^{(2)}= 0.4377(1)$, $\beta=0.25(2)$ from off-critical simulations and $\theta= 0.159(5)$ from time-dependent simulations, both are in agreement wit the DP exponents (see table \ref{table1}). We note that DP exponents were also obtained for other microscopic models \cite{ginelli03,lipowsky03}. Therefore, the observation of a first-order phase transition seems to be a transient effect.

Hinrichsen \cite{hinrichsenarxiv} argues about a possible connection between the pair contact process with diffusion (PCPD) and nonequilibrium wetting, pointing that the MN1 equation is equivalent to the Langevin equation for the PCPD \cite{howard97}. There has been a long discussion about the universality class of the PCPD \cite{henkel04}, and it is still not clear if the PCPD model is in the DP universality class or if it is in a new universality class of models with an absorbing state. The result obtained in \cite{munoz03} would be in agreement with the PCPD being in the DP universality class, however, the continuum approach may be inadequate for PCPD \cite{hinrichsenarxiv}.

Although the DP exponents were obtained in Fig. \ref{gMNDP}, the transition, that takes place with the parameters used there, is different from the transition at $p=0$. To see this one can consider the survival probability $P_s(t)$, which is the probability that the system will not enter the absorbing state until time $t$. At $p=0$ the survival probability can be less than one because once $\rho_0= 0$ it will remain zero in the dynamics that follows. It decays to zero, in a power-law (exponential) way at (above) criticality and it reaches a saturation value after some transient below criticality. For $p\neq 0$ the survival probability is always one (in the moving and in the bound phases) because even if the first-layer is completely filled evaporation events on it are  still possible. The Langevin equation related the RSOSW model at $p=0$ is the DP Langevin equation \cite{hinrichsen00}, where the noise term is multiplied by the square root of the field, and it cannot be obtained from the MN1 equation with a simple transformation of variables.          

The tricritical point of the bKPZ-- universality class was analyzed, with numerical integration of the MN1 equation and MC simulations of the SSW model, by Romera et al. \cite{romera08}. They obtained $\gamma= 0.35(2)$, $z=1.4(1)$, $\beta^{(2)}= 1.50(9)$, $\zeta^{(2)}= 0.9(1)$, $\theta^t= 0.49(2)$ and $\nu_\parallel^{t}= 2.0(2)$. As is the case of the bEW class, the critical exponents $\beta^t$, $\nu_\perp^t$ and $\nu_\parallel^t$ differ from the critical exponents of the bKPZ-- class while the scaling exponents are in agreement with the KPZ values $\gamma= 1/3$ and $z=3/2$. As far as we know, the critical behavior of the tricritical point of the bKPZ+ universality class was not yet analyzed.  

We point out that the wetting transition can be envisaged as a contact process with long-range interactions, considering the contact points with the substrate as active sites and the detached part between two contact points as generating an effective long range interaction between them \cite{ginelli05,ginelli06,hinrichsen07}. In the so-called $\sigma-$process \cite{ginelli05}, depending on the parameter controlling the long-range interactions the phase transition may be DP or first-order.   

\subsection{Extensions of the problem}

Hitherto only short range interactions between the wall and the absorbed particles were considered. In order to study long range interactions one can take the potential
\begin{equation}
V(h)= \frac{b}{sh^s}+\frac{c}{lh^l},
\label{potentiallong}
\end{equation}
where the parameters $b$ and $c$ have the same function as they have in the short-range interaction potential (\ref{2eqVh}) and $l>s$. Hammal et al. \cite{hammal06} studied complete wetting for the long range interactions case with numerical integration of the associated multiplicative noise equation, coming form the Cole-Hopf transformation of the bKPZ equation (\ref{bKPZ}) with the potential (\ref{potentiallong}),  and power counting arguments. The results are: for $s>1$ the critical behavior is the same as the one obtained with short range interactions for the bKPZ-- and bKPZ+ universality classes; for $s<1$ mean field (in the sense that the noise is irrelevant) critical behavior holds. These results for long range interactions were also confirmed with a microscopic model in \cite{hammal06}. Another microscopic model with long range interactions was previously studied in \cite{lipowsky03}, nevertheless, as pointed out in \cite{santos04}, it does not display a mechanism that produces surface tension (like the RSOS constraint for example) and, therefore, it is not clear if it corresponds to the bKPZ equation with the potential (\ref{potentiallong}).      

One relevant point in considering long range interactions is that it may play an important role in possible experimental realizations of nonequilibrium wetting. Another generalization of the problem, that can be central in an experiment, is to consider diffusion of single atoms in the interface. This generalization was studied in \cite{rossner06} using the RSOSW model with diffusion. In this new version of the model, another possible transition, taking place with rate $D$, is $h_i\to h_i-1$ and $h_j\to h_j+1$, where $j$ is one of the nearest neighbors of $i$, before the transition $h_i>h_{i\pm1}$ and the final configuration respects the RSOS condition. What was found in \cite{rossner06} is that, for $p\neq 1$ diffusion can shift the critical line but the critical behavior is still the same and at $p=1$ the critical behavior and the critical point do not change.   

Another extension, which is not yet studied, is nonequilibrium wetting with disorder. In equilibrium wetting, the random substrate and also random bulk cases were studied, and new physical properties are born from these situations \cite{dietrich86}. An open problem, in the equilibrium case, is what happens to critical wetting when a random substrate is considered \cite{forgacs88,derrida92,gangardt08}.

\section{Mean Field Approximations for microscopic models}
\label{5}
In this section we present mean field approximations for the RSOSW and SSW models. We first write down the master equation for surface growth models and then apply simple and pair mean field approaches in order to solve it. The first mean field approximation, for microscopic models for nonequilibrium wetting, was introduced in \cite{hinrichsen03a}. The presentation here follows the works of Ginelli and Hinrichsen \cite{ginelli04} for the SSW model and Barato and de Oliveira \cite{barato07} for the RSOSW model. We also obtain some new results concerning pair mean field approach for the RSOSW model.  

\subsection{Master equation}
A stochastic process with continuous time evolves according to the master equation (see 
\cite{vankampen}), such equation reads 
\begin{equation}
\frac{d}{dt}P_\sigma(t)= \sum_{\sigma'\neq\sigma}\bigg(P_{\sigma'}(t)w_{\sigma'\to\sigma}-P_{\sigma}(t)w_{\sigma\to\sigma'}\bigg),
\end{equation}
where $P_\sigma(t)$ is the probability of being in a configuration $\sigma$ at time $t$ and $w_{\sigma'\to\sigma}$ is the transition rate  from the configuration $\sigma'$ to the configuration $\sigma$. In the case of surface growth models with  deposition and evaporation rates depending only on height of the target site and it's nearest neighbors, the master equation becomes
\begin{eqnarray}
\frac{d}{dt}P(h_1,h_2,...,t)= \sum_{n}\sum_{i=1}^{L}\{ w_n(h_{i-1},h_i-n,{h_i+1})P(h_1,...,h_i-n,...,t)\nonumber\\
-w_n(h_{i-1},h_i,h_{i+1})P(h_1,...,h_i,...,t)\},
\label{4master}
\end{eqnarray} 
where $w_n(h_{i-1},h_i-n,h_{i+1})$ is the rate for a transition from $(...,h_{i-1},h_i-n,h_{i+1},...)$ to
$(...,h_{i-1},h_i,h_{i+1},...)$, $L$ is the system size and we are considering a one-dimensional system with periodic boundary conditions. 

Denoting $w_n(h_{i-1}=k,h_i=l,h_{i+1}=m)$ by $w_n(k,l,m)$ we have that for the RSOSW model, where the height changes by only $n=\pm1$, the rates are:
\begin{equation}
w_{+}(k,k,k)= w_+(k+1,k,k)= w_+(k,k,k+1)= w_+(k+1,k,k+1)= q,
\end{equation}
\begin{equation}
w_{-}(k,k+1,k)= w_-(k,k+1,k+1)= w_-(k+1,k+1,k)= 1,
\end{equation}
and
\begin{equation}
w_{-}(k+1,k+1,k+1)= p.
\end{equation}
On the other hand, for the SSW model they are non-zero only for $n=\pm2$ and given by   
\begin{equation}
w_{+2}(k,k-1,k)= p
\end{equation}
and
\begin{equation}
w_{-2}(k,k+1,k)= 1-p.
\end{equation}

In the following we denote $P(k,l,m,t)w_n(k,l,m)$ by $J_n(k,l,m,t)$. The time evolution of the one-site probability distribution is obtained by summing over all heights but one in equation (\ref{4master}), which gives
\begin{equation}
\frac{d}{dt} P(l,t) = \sum_{k,m}\sum_{n}\{ J_n(k,l-n,m,t) -J_n(k,l,m,t) \},
\label{4master1}
\end{equation} 
For the the pair mean field approach (see below) we also need the time evolution of the two-site probability distribution, it is given by
\begin{eqnarray}
\frac{d}{dt} P(k,l,t)= \sum_{m} \sum_{n}\{ J_n(m,k-n,l,t)+J_n(k,l-n,m,t)\nonumber\\
-J_n(m,k,l,t)- J_n(k,l,m,t) \}.
\label{4master2}
\end{eqnarray}

The scaling exponents of the KPZ and EW universality classes are related to the interface width $w$. One can make the stronger assumption that the one-site probability distribution of these universality classes, for an infinite system and in the long time limit, is given by 
\begin{equation}
P(h,t)= t^{-\gamma}f\bigg(\frac{h-vt}{t^\gamma}\bigg),
\label{4ansatz}
\end{equation}
where $f(x)$ is a scaling function. Since in the calculations that follows we always consider an infinite system in the long time limit, when possible to solve the problem exactly, we use this ansatz for the one-site probability distribution at the wetting transition and in the moving phase. 

The mean field approximations we apply to the RSOSW and SSW models in the following consist in approximating the probability distribution $P(h_1,h_2,...,t)$ by a (simple or pair) factorized form so that the master equation becomes tractable. They are applied to one-dimensional models and expected to capture some features of them, they are not expected to become valid above some critical dimension.

\subsection{The RSOSW model}

\subsubsection{Simple mean field approximation}

In simple mean field, terms like $P(k,l,m,t)$ are approximated by their factorized form, i.e.,
\begin{equation}
P(k,l,m,t)= P(k,t)P(l,t)P(m,t).
\end{equation}
Note that this approximation does not take the RSOS constraint into account. For example, the probability $P(k,k+2,k,t)$ is zero in the original problem and in the simple mean field approach it is simple given by $P(k,k+2,k,t)= P(k,t)^2P(k+2,t)$.

With this approach the master equation for the one site probability distribution (\ref{4master1}) for the RSOSW model acquires the following simpler form,
\begin{eqnarray}
\frac{d}{dt}P_k= q(P_{k-1}^3-P_{k}^3)+(1-2q)(P_{k}^2P_{k+1}-P_{k-1}^2P_{k})+\nonumber\\
(2-q)(P_{k}P_{k+1}^2-P_{k-1}P_{k}^2)+p(P_{k+1}^3-(1-\delta_{k,0})P_{k}^3),
\label{4mastermf1}
\end{eqnarray}
where $P_k$ denotes the one-site probability distribution, $k\ge0$ and $P_{-1}=0$ (in order to account for the presence of the wall at height zero). The factor $(1-\delta_{k,0})$, multiplying $P_k^3$, comes from the fact that evaporation at zero height is forbidden. The initial condition, corresponding to an initially flat interface, is $P_k=\delta_{k,0}$. 

Below criticality we assume that $P_k$ decays exponentially, 
\begin{equation}
P_k= Ax^k,
\label{4expansatz}
\end{equation}
where $A=1-x$ is a normalization constant and $x<1$. This ansatz is valid below the critical line where the interface is bounded, while at criticality $x\to 1$. By substituting (\ref{4expansatz}) in equation (\ref{4mastermf1}) we obtain
\begin{equation}
-px^3+(q-2)x^2+(2q-1)x+q=0,
\label{4xmf1}
\end{equation}
which, with $x=1$, gives the critical line
\begin{equation}
q_c= \frac{1}{4}(p+3).
\label{4criticalmf1}
\end{equation}
With the probability distribution (\ref{4expansatz}) we have $P_0= 1-x$, $\langle h\rangle= \frac{x}{1-x}$ and $w= \frac{\sqrt{x}}{1-x}$. From equations (\ref{4xmf1}) and (\ref{4criticalmf1}), one can verify that near the critical line 
\begin{equation}
x\approx 1-\frac{2(q_c-q)}{2}.
\end{equation} 
Therefore,
\begin{equation}
P_0\sim (q_c-q)^1,\qquad \langle h\rangle\sim(q_c-q)^{-1},\qquad w\sim (q_c-q)^{-1},
\end{equation}
giving $\beta=1$ and $\zeta=1$.

In order to solve (\ref{4mastermf1}) at and above the critical line we take the continuum limit, where $P_k(t)\to P(h,t)$ and $h=k\delta$. In this limit, to second order in $\delta$, equation (\ref{4mastermf1}) becomes
\begin{equation}
\partial_tP(h,t)= -\delta12(q-q_c)P^2\partial_hP+\delta^2(2q+1+3p)\bigg(P(\partial_hP)^2+\frac{1}{2}P^2\partial^2_hP\bigg).
\label{RSOSMFcont}
\end{equation}
In the long time limit higher order terms are irrelevant even with $\delta=1$. To see this one can carry out the calculation that follows with general $\delta$ and then verify it with the final result. Since we are interested in this limit we proceed with the calculations setting $\delta=1$.

At the critical point the first term on the right hand side of the above equation vanishes, using the ansatz (\ref{4ansatz}) we obtain consistency only if $\gamma=1/4$ and $v=0$. The differential equation for the scaling function is
\begin{equation}
f(x)+xf'(x)+2(2q+1+3p)[2f(x)f'(x)^2+f(x)^2f''(x)]=0,
\end{equation}
which has the solution $f(x)=\sqrt{1-\frac{x^2}{2(2q+1+3p)}}$, giving 
\begin{equation}
P(h,t)= \sqrt{t^{-1/2}-\frac{h^2}{2t(2q+1+3p)}}.
\end{equation}

Above the critical point only the first term in the right hand side of equation (\ref{RSOSMFcont}) matters, the second is irrelevant for $t\to \infty$. By following the same procedure as in the previous case, we obtain $\gamma=1/3$, $v=0$ and
\begin{equation}
f(x)+xf'(x)-36(q-q_c)f(x)^2f'(x)=0,
\end{equation}
which has the solution $f(x)= \sqrt{x/12(q-q_c)}$, giving
\begin{equation}
P(h,t)= \sqrt{\frac{h}{12(q-q_c)t}},
\end{equation}
for $q>q_c$.

Very surprising is the fact that with this very simple approximation we obtain the KPZ growth exponent $\gamma= 1/3$ above the critical line. A problem with it is that we do not obtain the mean height growing linearly with time above criticality, i.e., $v=0$ also for $q>q_c$. The next step is to perform a improved approximation, that satisfies the RSOS condition and gives $v>0$ in the moving phase.

\subsubsection{Pair mean field approximation}

In the pair mean field approximation $P(k,l,m,t)= P(k,l,t)P(l,m,t)/P(l,t)$. This means that the probability distribution is approximated by a pair factorized form. Clearly, the present approach satisfies the RSOS condition.

\begin{figure}
\begin{center}
\includegraphics[width=85mm]{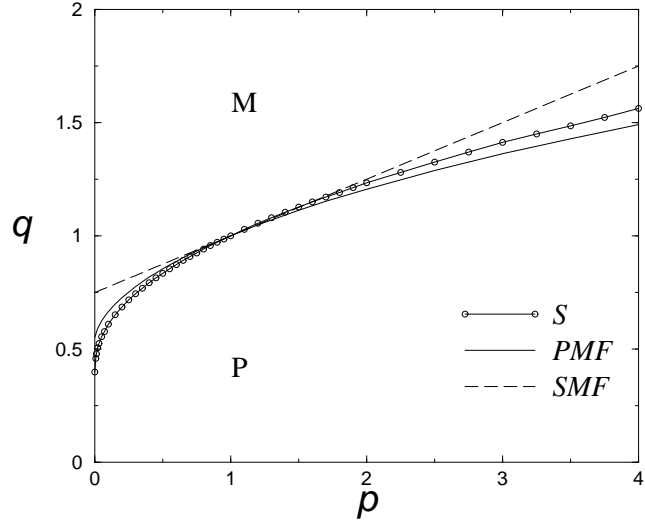}
\caption{Comparison between the phase  diagrams obtained with mean filed approximations, simple (SMF) and pair (PMF), and simulations (S). Figure taken from \cite{barato07}.} 
\label{gpairMFresults}
\end{center}
\end{figure}

For the pair mean field we need the time evolution of one-site and two-site probability distributions. Because of the RSOS condition only $P_{k,k}$, $P_{k,k-1}$ and $P_{k,k+1}$ are non-zero. Therefore, we have four variables: the three two-site probability distributions and the one-site probability distribution $P_k$. They are not all independent, one obvious constraint is that $P_k= P_{k,k}+ P_{k,k+1}+ P_{k,k-1}$ and  the other is $P_{k,k+1}= P_{k+1,k}$. The second comes from the fact that the transition rates are symmetric and initially the number of up steps is equal to the number o down steps (flat interface). Therefore, we are left with two independent set of equations. 

Applying the pair mean field to equation (\ref{4master2}) we obtain
\begin{eqnarray}
\frac{d}{dt}x_k= 2q\biggl[\frac{y_{k-1}(y_{k-1}+x_{k-1})}{P_{k-1}}-\frac{x_k(x_k+y_k)}{P_k}\biggr]\nonumber\\
+2\biggl[\frac{y_k(y_k+x_{k+1})}{P_{k+1}}-\frac{x_ky_{k-1}}{P_k}\biggr]-2p\frac{x_k^2}{P_k}(1-\delta_{k,0}),
\label{eqxk}
\end{eqnarray}
\begin{equation}
\frac{d}{dt}y_k= q\frac{x_k^2-y_k^2}{P_k}- \frac{y_k^2}{P_{k+1}} + p\frac{x_{k+1}^2}{P_{k+1}},
\label{eqyk}
\end{equation}
where $x_k= P_{k,k}$ and $y_k= P_{k+1,k}= P_{k,k+1}$. The term $1-\delta_{k,0}$ multiplying $2p\frac{x_k^2}{P_k}$ in the first equation comes from the fact that evaporation events are forbidden at height zero and the initial condition is $x_k= \delta_{k,0}$ and $y_k=0$ for all $k$.

The above equations were integrated numerically in \cite{barato07}, the results are in agreement with $\beta=1$, $\zeta=1/3$, the EW growth exponent $\gamma=1/4$ at the critical line and  the KPZ growth exponent $\gamma= 1/3$ above it. Also, with this improved approximation an interface growing linearly with time for $q>q_c$ is observed. The phase diagrams, coming from mean field, are compared to the one obtained with simulations in Fig. (\ref{gpairMFresults}). One can see that the agreement of pair mean field is much better when compared to simple mean field.  

In relation to the simple mean field the main improvements are: better agreement with simulations, when comparing the phase diagram and also observables (not showed here see \cite{barato07}), and a interface growing with a non-zero velocity in the moving phase. The exponent $\zeta=1/3$ is different from the one obtained with simple mean field, which is $\zeta=1$, while the others are the same (including the EW and KPZ growth exponents at and above criticality respectively). 

An interesting feature of the pair mean field is that it becomes the exact solution of the model when $p=1$. From equation (\ref{eqDB2}), with the pair mean field approach, follows that
\begin{equation}
q\frac{x_k}{P_k}= \frac{x_{k+1}}{P_{k+1}},\qquad y_k^2=x_kx_{k+1},
\label{pairMFexact1}
\end{equation} 
which has the solution
\begin{equation}
\frac{x_k}{P_k}= \Lambda^{-1}q^k,\qquad \frac{y_k}{\sqrt{P_kP_{k+1}}}= \Lambda^{-1}q^{q+1/2}.
\label{pairMFexact2}
\end{equation} 
Equations (\ref{pairMFexact1}) and (\ref{pairMFexact2}), with the condition $P_k= x_k+y_k+y_{k-1}$, give
\begin{equation}
q^{k-1/2}\sqrt{P_{k-1}}+q^{k}\sqrt{P_{k}}+q^{k+1/2}\sqrt{P_{k+1}}= \Lambda\sqrt{P_k}.
\end{equation} 
This is equal to equation (\ref{eq3eigen2}), with the components of the eigenvector, associated to the maximum eigenvalue $\Lambda$, given by $\phi_k=\sqrt{P_k}$. 

The pair mean field becomes the exact solution because, at $p=1$ and $q\le 1$, the probability distribution is pair-factorized in the stationary sate. This comes from the fact that the stationary probability distribution can be written as a product of transfer matrices. A natural question that arises is whether the time-dependent probability distribution is also pair-factorized. If this is the case, the solution of the pair mean field equations (\ref{eqxk}) and (\ref{eqyk}) would give the exact time-dependent probability distribution. In order to check this, we compare the one-site probability distribution obtained from simulations and numerical integration of (\ref{eqxk}) and (\ref{eqyk}) in Fig. \ref{gpairMFresults}. We see that they are in agreement, suggesting that the time-dependent probability distribution is indeed pair-factorized.     

\begin{figure}
\begin{center}
\includegraphics[width=95mm]{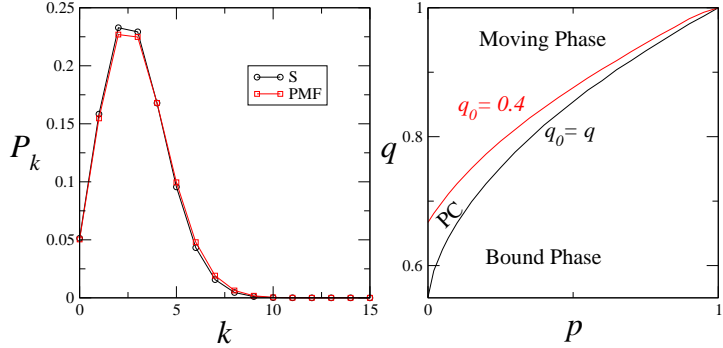}
\caption{Pair mean field results. Left: comparison of the one site probability distribution obtained from pair mean field (PMF) and from simulations (S) at $q= 0.99$ and $t=500$. Right: phase coexistence (PC) region at the $q\times p$ plane with $q_0= 0.4$ obtained from numerical integration of the pair mean field equations.} 
\label{gpairMFresults}
\end{center}
\end{figure}

The case $q_0\neq q$ can also be considered within the pair mean field approach. Equations (\ref{eqxk}) and (\ref{eqyk}), for $k=0$ and $k=1$, have to be modified to
\begin{equation}
\frac{d}{dt}x_0= -2q_0\frac{x_0(x_0+y_0)}{P_0}+ \textrm{evaporation part}
\end{equation}
\begin{equation}
\frac{d}{dt}x_1= 2q_0\frac{y_{0}(y_{0}+x_{0})}{P_{0}}-2q\frac{x_1(x_1+y_1)}{P_1}+ \textrm{evaporation part}
\end{equation}
\begin{equation}
\frac{d}{dt}y_0= q_0\frac{x_0^2-y_0^2}{P_0}+ \textrm{evaporation part},
\end{equation}
where the evaporation part corresponds to the terms of equation (\ref{eqxk}) and (\ref{eqyk}) that are not multiplied by $q$. By integrating these new equations we obtain: for $q_0<q_0^*$ the transition becomes first-order for $p\ge1$ while for $p<1$ a phase coexistence region (in the sense explained in Sec. \ref{2}) is observed, and the transition at the additional line $q_c^{(2)}> q_c$ seems to be first-order. In Fig. \ref{gpairMFresults} we show the phase coexistence region in the $q\times p$ plane for $q_0= 0.4$ obtained with numerical integration of the above equations. The results presented in Fig. \ref{gpairMFresults} are new.

\subsection{The SSW model}

For the SSW model, with the pair mean field approach, we are left just with one independent set of equations. This happens because, in comparison to the RSOSW model, there are the same two constraints and not four but three variables ($P_k$, $P_{k,k+1}$ and $P_{k,k-1}$). We can define the height for an interface, satisfying the single-step constraint, in the following way. With every pair of sites with heights given by $h_i,h_{i+1}$, we associate a variable $h_i'= \textrm{min}(h_i,h_{i+1})$, that can be viewed as defined in a point between the two sites. For example, the initial configuration, where $h_i=0$ ($h_i=1$) if $i$ is even (odd), corresponds to $h_i'=0$ for all $i$. In this way the two-site probability distribution $P_{k,k+1}$ can be viewed as a one-site probability distribution $P_{j}$, where $j=k$. In the calculations that follows we use this new one-site probability distribution.

The master equation for the (new) one-site probability distribution within pair mean field for the SSW model reads
\begin{eqnarray}
\frac{d}{dt}P_j= p\bigg(\frac{P_{j-1}^2}{P_{j-1}+P_{j-2}}-\frac{P_{j}^2}{P_{j}+P_{j-1}}\bigg)\nonumber\\
+(1-p)\bigg(\frac{P_{j+1}^2}{P_{j+1}+ P_{j+2}}-\frac{P_{j}^2}{P_{j}+ P_{j+1}}\bigg),
\label{MFSS}
\end{eqnarray}
where the boundary conditions, to be specified below, depend on the velocity of the wall.

The above equation in the continuum limit reads
\begin{eqnarray}
\partial_tP(h,t)= -\delta(p-1/2)\partial_hP+
\delta^3\frac{(p-1/2)}{12}\bigg(3\frac{(\partial_hP)^3}{P^2}-6\frac{\partial_hP\partial_h^2P}{P}+4\partial_h^3P\bigg)+\nonumber\\
\delta^4\bigg(\frac{(\partial_hP)^4}{4P^3}-5\frac{(\partial_hP)^2\partial_h^2P}{8P^2}+
\frac{(\partial_h^2P)^2}{4P}+\frac{\partial_hP\partial_h^3P}{4P}-\frac{\partial_h^4P}{8}\bigg)
\label{MFSSconti}
\end{eqnarray}
where $h= j\delta$ and we went until order $\delta^4$. As we did in the simple mean field for the RSOSW model we set $\delta=1$, because one can show that higher order terms are  irrelevant in the long time limit \cite{ginelli04}.

In order to solve equation (\ref{MFSSconti}) we apply the ansatz (\ref{4ansatz}) to it. The first term produces a linear propagation of the interface, therefore   
\begin{equation}
v= p-1/2,
\end{equation}
which is in agreement with equation (\ref{2veloSS}). At $p=1/2$ the first two terms on the right hand side of (\ref{MFSSconti}) vanish and a non-trivial equation is obtained only if $\gamma=1/4$, giving the EW growth exponent in the equilibrium case. For $p\neq 1/2$ the third term on the right hand side of (\ref{MFSSconti}) is irrelevant in the long time limit and a non-trivial equation is obtained only if $\gamma=1/3$, giving the KPZ growth exponent.

We proceed, presenting the resulting differential equations for the scaling function and their solutions for the cases $p=1/2$, $p=0$ and $p=1$. In each case subtle boundary conditions have to be used, in order to account for the moving substrate. Our aim is to calculate the exponent $\theta$, within this approximation, for the bEW and bKPZ universality classes. To avoid confusion we denote the scaling function $f$ and the critical exponent $\theta$, for each value of $p$, by $f_p$ and $\theta_p$.           

\subsubsection{bEW case}

At $p=1/2$ the differential equation for the scaling function $f_{1/2}(x)$ is 
\begin{eqnarray}
(f_{1/2})^{-3}\bigg[(f_{1/2})^4-\frac{5}{2}f_{1/2}(f_{1/2}')^2f_{1/2}''+(f_{1/2})^2\bigg((f_{1/2}'')^2+f_{1/2}'f_{1/2}'''\bigg)\nonumber\\
+(f_{1/2})^3\bigg(xf_{1/2}'-\frac{1}{2}f_{1/2}'''\bigg)+(f_{1/2}')^4\bigg]=0,
\label{MFSSEW}
\end{eqnarray}
where $x= ht^{-\gamma}$. Integrating it we get
\begin{equation}
2xf_{1/2}-\frac{(f_{1/2}')^3}{(f_{1/2})^2}+2\frac{f_{1/2}'f_{1/2}''}{f_{1/2}}-f_{1/2}'''=0.
\label{MFSSEWint}
\end{equation}
We have to solve the above equation with the appropriate boundary conditions. At $p=1/2$ the substrate is fixed at height zero, therefore, the evolution of the probability distribution $P_j$ follows equation (\ref{MFSS}) for $j>0$, while at zero height it follows
\begin{eqnarray}
\frac{d}{dt}P_0= -\frac{1}{2}P_{0}+ \frac{1}{2}\bigg(\frac{P_{1}^2}{P_{1}+ P_{2}}\bigg),
\end{eqnarray}
where the missing terms in the equation come from the facts that evaporation is forbidden at the substrate and $P_j= 0$ for $j<0$. Now if we assume that the scaling function satisfies $f'(0)<\infty$ we have that $P_j\approx t^{-1/4}f(0)$ for $j=0,1,2$. Substituting this in the last equation, we see that it is valid only if $f(0)=0$ (the left side of the equation is proportional to $t^{-5/4}$ and the right side is proportional to $t^{-1/4}$).

The solution of equation (\ref{MFSSEWint}), satisfying the boundary condition $f(0)=0$, is \cite{ginelli04}
\begin{equation}
f_{1/2}(x)= \frac{2^{5/4}}{\sqrt{\pi}}x^2\exp(-x^2/\sqrt{2}),
\end{equation} 
which gives $\theta_{1/2}= 3/4$.

\subsubsection{bKPZ case}

For $p\neq 1/2$ the differential equation we get is  
\begin{eqnarray}
(f_p)^{-2}\bigg[8(f_p)^3+6v(f_p')^3-12vf_pf_p'f_p''+8(f_p)^2(xf_p'+vf_p''')\bigg]=0,
\label{SSMFKPZ}
\end{eqnarray}
where $x= (h-vt)t^{-\gamma}$. 

Obtaining the suitable boundary conditions, for the cases $p=0$ and $p=1$, that accounts for a moving wall, is more involved. With an argument similar to the one presented above it is possible to show that $f_1(0)= 0$ and $f_0(0)\neq 0$  \cite{ginelli04}. With these boundary conditions the solutions of equation (\ref{SSMFKPZ}), that are physically suitable, are \cite{ginelli04}: 
\begin{equation}
p=1:\qquad f_1(x)\propto \left\{\begin{array}{cc}
\bigg[Ai\bigg(\frac{-x}{2^{1/3}}\bigg)- 3^{-1/2}Bi(\frac{-x}{2^{1/3}})\bigg]^4 & \textrm{if $0\le x<x_0$}\\
0 & \textrm{if $x_0\le x<\infty$}
\end{array}\right.
\label{f_1(x)}
\end{equation}   
and 
\begin{equation}
p=0:\qquad f_0(x)\propto Ai\bigg(\frac{x}{2^{1/3}}\bigg)^4,
\label{f_0(x)}
\end{equation}
where $Ai(x)$ and $Bi(x)$ are Airy functions and $x_0\approx 3.32426$. The scaling function (\ref{f_1(x)}) gives the exponent $\theta_1= 4/3$ for the bKPZ-- case and (\ref{f_0(x)}) $\theta_0= 1/3$ for the bKPZ+ case. They are different from of the exponents obtained from numerical simulations (see table \ref{table1}).   
   
\begin{figure}
\begin{center}
\includegraphics[width=95mm]{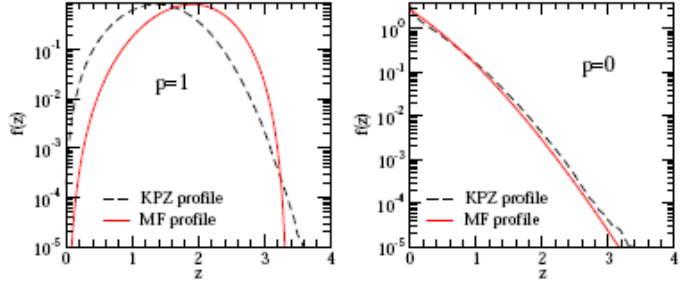}
\caption{The scaling function $f(z)$ for the bKPZ-- ($p=1$) and the bKPZ+ ($p=0$) universality classes, obtained from numerical simulations (doted line) and mean field (MF) approach for the SSW model. Figure taken from \cite{ginelli04}.} 
\label{MFSSWresults}
\end{center}
\end{figure}

Although the mean field theory does not predict the correct value of the critical exponent $\theta$ in the nonequilibrium cases, it does predict differences between the bKPZ+ and bKPZ-- universality classes. For the bKPZ-- (bKPZ+) an exponent bigger (smaller) than the equilibrium exponent $\theta_{1/2}= 3/4$ is obtained. Also, as is shown in Fig \ref{MFSSWresults}, the scaling functions, given in equations (\ref{f_1(x)}) and (\ref{f_0(x)}), are qualitatively similar to the scaling functions obtained from numerical simulations.      

With this mean field approach the one-site probability distribution for the free interface case can also be obtained, in this case the differential equations for the scaling functions have to be solved with different boundary conditions \cite{ginelli04}. In \cite{prahofer1,prahofer2,prahofer3,prahofer4}, several exact results were obtained for the free interface case with the polynuclear growth model, which is in the KPZ universality class. A very important question is whether, with the methods used in \cite{prahofer1,prahofer2,prahofer3,prahofer4}, the bounded interface case can also be treated analytically, allowing one to compute the exponent $\theta$ for the bKPZ universality classes exactly.

\section{Nonequilibrium wetting in higher dimensions}
\label{6}
We now turn to nonequilibrium wetting in higher dimensions. First we present a mean field approach to the MN equations, which yields a qualitatively picture of what happens above the critical dimension. Then we discuss the critical behavior that is expected in higher dimensions using power counting and renormalization group arguments. We note that the description we presented for the one-dimensional case is much more complete and, therefore, if one wants a better understanding on what follows one should turn to the references presented below. 

\subsection{Mean field for the continuum model}

The mean field approximation to the MN equations is done in the following way \cite{genovese99}. The first step is to approach the discrete laplacian by 
\begin{equation}
\nabla^2n_i= \frac{1}{2d}\sum_j(n_j-n_i)\approx \langle n\rangle- n_i,
\label{approxlaplacian}
\end{equation} 
where the sum runs over all nearest neighbors and this approximation is expected to become valid above some critical dimension. The equation that results from it, in the MN1 case, is
\begin{equation}
\frac{d}{dt}n= -an-bn^{s+1}-cn^{2s+1}+\sigma(n-\langle n\rangle)+ n\zeta.
\end{equation}
The associated Fokker-Planck equation \cite{gardiner,vankampen} reads
\begin{eqnarray}
\frac{\partial}{\partial t}P(n,\langle n\rangle, t)= -\frac{\partial}{\partial n}\{[-(a-D/2)n-bn^{s+1}-cn^{2s+1}+\sigma(n-\langle n\rangle)]P(n,\langle n \rangle,t)\}\nonumber\\+\frac{D}{2}\frac{\partial^2}{\partial n^2}[n^2P(n,\langle n\rangle,t)].
\end{eqnarray}
Solving the above equation in the stationary state we obtain
\begin{equation}
P_s(n,\langle n\rangle)\propto n^{2(a-\sigma)/D-1}\exp\bigg(-\frac{2b}{D s}n^s-\frac{c}{D s}n^{2s}-\frac{2\sigma\langle n\rangle}{nD}\bigg),
\label{PsMFMN1}
\end{equation}
where $P_s(n,\langle n\rangle)$ is the probability distribution in the stationary sate. Finally the critical behavior of the order parameter $\langle n\rangle$ is obtained with the self-consistency condition
\begin{equation}
\langle n\rangle= \frac{\int_0^\infty nP_s(n,\langle n\rangle)dn}{\int_0^\infty P_s(n,\langle n\rangle)dn}.
\label{self}
\end{equation}
For the MN2 case we consider the non-order parameter Langevin equation (\ref{nonorder}), with the approximation (\ref{approxlaplacian}). The solution of the corresponding Fokker-Planck equation is  
\begin{equation}
P_s(m,\langle m\rangle)\propto m^{2(a-\sigma)/D-1}\exp\bigg(-\frac{2b}{D s}m^{-s}-\frac{c}{D s}m^{-2s}-\frac{2\sigma\langle m\rangle}{mD}\bigg).
\label{PsMFMN2}
\end{equation}

With the mean field approach introduced we proceed considering complete wetting, therefore, we set $c=0$ and $b$ constant and positive. In the calculations below, for the MN2 case we follow \cite{hammal05} and for the MN1 case we follow \cite{birner02,munoz05}.

For the MN2 case, the self-consistency condition (\ref{self}) with the probability distribution (\ref{PsMFMN2}) leads to the following equation,
\begin{equation}
\langle m\rangle= \frac{I_1(\langle m\rangle)}{I_0(\langle m\rangle)},
\label{selfMN2}
\end{equation}
where 
\begin{equation}
I_k(y)= \int_0^\infty dx x^{k+2(a-\sigma)/D-1}\exp\bigg(-\frac{2b}{D s}x^{-s}-\frac{2\sigma y}{xD}\bigg).
\end{equation}
We want to calculate the exponent $\beta$, defined by $\langle m\rangle\sim (a_c-a)^{-\beta}$ ($m$ is a non-order parameter). We can write $I_k(y)$ as
\begin{equation}
I_k(y)= \bigg(\frac{2\sigma y}{D}\bigg)^{k+2(a-\sigma)/D}\int_0^{\infty}dz
z^{-1-k-2(a-\sigma)/D} e^{-z}\exp\bigg[-\frac{2b}{Ds}\bigg(\frac{D z}{2\sigma y}\bigg)^s\bigg]  
\end{equation}  
where $z= \frac{2\sigma y}{D x}$. At criticality $\langle m\rangle\to \infty$, we are interested in the asymptotic form of $I_k(y)$, with $y\to \infty$. In this limit
\begin{eqnarray}
I_k(y)\approx \bigg(\frac{2\sigma y}{D}\bigg)^{k+2(a-\sigma)/D} \Gamma(-2a/D+2\sigma/D-k)\nonumber\\
-\bigg(\frac{2\sigma y}{D}\bigg)^{k+2(a-\sigma)/D-s} \frac{2b}{Ds} \Gamma(-2a/D+2\sigma/D-k+s),
\end{eqnarray} 
where the term $\exp\bigg[-\frac{2b}{Ds}\bigg(\frac{D z}{2\sigma y}\bigg)^s\bigg]$ was expanded until first order in $(z/y)^s$. The above asymptotic form for $I_k(y)$ and equation (\ref{selfMN2}) lead to \cite{hammal05}
\begin{equation}
\langle m\rangle \sim (-D/2-a)^{-1/s},
\end{equation}
which gives $\beta= 1/s$ and $a_c= -D/2$. 

For the MN1 case the resulting equation, coming from (\ref{PsMFMN1}) and (\ref{self}), is \cite{birner02,munoz05}
\begin{equation}
\langle n\rangle= \frac{J_1(\langle n\rangle)}{J_0(\langle n\rangle)},
\end{equation}
where
\begin{equation}
J_k(y)= \int_0^\infty dx x^{k+2(a-\sigma)/D-1}\exp\bigg(-\frac{2b}{D s}x^{s}-\frac{2\sigma y}{xD}\bigg).
\end{equation}
It can be shown that, for $y\to 0$, \cite{birner02,munoz05}
\begin{equation}
J_k(y)\approx A_k+ B_k y^{2(a-\sigma)/D+k}+ C_k y^{2(a-\sigma)/D+k+s},
\label{Ikassym}
\end{equation}
leading to
\begin{equation}
\langle n\rangle\sim (D/2-a)^{\textrm{max}[1/s,D/2\sigma]}
\end{equation}
where max indicates the maximum. This shows that for the MN1 case we have a more complex critical behavior: a weak-noise regime, where $D<2\sigma/s$ and $\beta= 1/s$, and a strong-noise regime, where $D>2\sigma/s$ and $\beta= D/2\sigma$. With the asymptotic form (\ref{Ikassym}) one can also calculate the critical exponents related to higher order moments $\langle n^k\rangle= J_k(\langle n\rangle)/J_{k-1}(\langle n\rangle)$. The strong-noise regime can be further divided into two regimes that are different with respect to the critical behavior of higher order moments \cite{munoz05}.   

Critical wetting can be studied with the same kind of procedure, this was done by de los Santos et al. \cite{santos07}. In critical wetting, for the MN2 case just one regime is found, with the critical exponent being independent of the noise strength,  while for the MN1 case weak and strong noise regimes are found. Moreover, with the present mean field approximation, phase coexistence for the bKPZ-- universality class, when the attraction between the substrate and the interface is strong enough, is observed \cite{giada00,santos03}.             

We point out that in contrast to the method presented above a simpler approach would be consider the one-variable case by taking out the Laplacian. The one-variable MN1 equation was solved exactly in \cite{graham82}, it has a rich scaling behavior, nevertheless it does not display the strong noise regime. One interesting feature of the one-variable approximation is that it allows one to clarify essential differences between the MN1 and the DP Langevin equations \cite{munoz98}.

\subsection{Scaling analysis}

Given a Langevin equation we can define the partition function $Z$ by summing over all configurations and realizations of noise that satisfy it. In the case of the MN1 equation (\ref{MN1}) with the potential (\ref{2eqVn}) and $c=0$ it reads 
\begin{equation}
Z\propto \int Dn D\zeta P[\zeta]\delta\bigg(\frac{\partial}{\partial t}n+ an+ bn^{s}-\nabla^2n-n\zeta\bigg),
\end{equation}
where 
\begin{equation}
P[\zeta]\propto \exp(\zeta^2/2D)
\end{equation} 
and $\int Dn D\zeta$ denotes a functional integration. With the introduction of a response field $\tilde{n}$ one can integrate out the noise \cite{janssen76, janssen81}, resulting in the following equation,
\begin{equation}
Z\propto \int Dn D\tilde{n} \exp\bigg(-S[n,\tilde{n}]\bigg),
\end{equation}
with the action $S[n,\tilde{n}]$ given by
\begin{equation}
S[n,\tilde{n}]= \int d^dxdt\bigg[\frac{D}{2}\tilde{n}^2n^2-\tilde{n}\bigg(\frac{\partial}{\partial t}n+an+bn^p-\nabla^2n\bigg)
\bigg],
\label{action}
\end{equation}
where, for simplicity, we set $\sigma=1$. Preforming naive power counting in this action we obtain
\begin{equation}
[n]+[\tilde{n}]=-d\qquad \textrm{and}\qquad [D]= d-2,
\end{equation}
where $[x]$ represents the dimension of the quantity $x$ in units of length. Therefore, the critical dimension, above which the noise becomes irrelevant, is $d_c=2$. It is known that the response field $\tilde{n}$ scales as the survival probability \cite{munoz97}. In the case of the bKPZ-- universality class the survival probability is always one, therefore the dimension of the response field $\tilde{n}$ is zero, giving $[n]= d$. Power counting at the critical dimension $d_c=2$ gives
\begin{equation}
\beta=1,\qquad \nu_\perp= 1/2,\qquad\nu_\parallel=1.
\label{mfexponents}
\end{equation}

Using standard methods \cite{zinn-justin} one can perform a perturbative expansion with the action (\ref{action}) and then obtain the renormalization group flow diagram for the MN1 equation. The calculations can be found  in \cite{munoz04}, in the following we discuss some important points of it. Since the term proportional to $b$ goes to zero in the moving phase, the bKPZ equation is equivalent to the KPZ equation in the moving phase, the flow diagram is similar to the well-known KPZ one \cite{wiese98}: above the critical dimension $d_c=2$ there is a weak-noise attractive fixed point, where $D=0$, and a strong-noise repulsive fixed point. If the noise strength is smaller then a certain threshold the flow runs to the weak noise fixed point and the critical exponents are given by (\ref{mfexponents}) if it is larger the flow runs to infinity and the critical behavior is not accessible through perturbation theory. Therefore the situation is similar to the one obtained with the mean field approximation. This was verified with numerical integration of the MN1 equation for $d=3$ in \cite{genovese99}. 

Another important point about the perturbative expansion is that changing $s$, the hardness of the wall,  does not introduce new divergences and therefore it is not expected to affect the critical behavior \cite{munoz04}. This is in agreement with numerical results and different from the mean field result that predicts an exponent depending on $s$ in the weak noise regime.  

The critical behavior at the critical dimension $d=2$ cannot be determined by perturbation theory. This is a very important case because in experimental situations the substrate is usually two-dimensional. The critical exponents $\theta$ (or $\beta$) of the bKPZ universality classes in two dimensions are not known. 

The renormalization group flow equations can be derived in an alternative way without writing down the action, but direct from the Langevin equation. For the MN1 equation this can be found in \cite{grinstein96} and for the KPZ equation in \cite{barabasi95}.

\section{Final remarks}
\label{7}
Just as the KPZ equation represents a robust universality class of nonequilibrium growing free interfaces, the bKPZ equation is expected to represent a robust universality class of nonequilibrium growing interfaces in the presence of wall. While equilibrium wetting transitions can be studied with the bEW equation, nonequilibrium wetting transitions are described by the bKPZ equation. Below we point out what we consider the main open problems in nonequilibrium wetting.  

With the introduction of the wall critical exponents arise, just one of them is independent while the others can be determined from scaling relations and the KPZ scaling exponents. An important open problem in nonequilibrium wetting is the exact calculation of the exponent $\theta$ for the bKPZ universality classes. While an exact solution for the free interface case, for a specific microscopic model in the KPZ universality class, is known \cite{prahofer1,prahofer2,prahofer3,prahofer4} the bounded case still remains without an exact solution. As we showed here within an mean field approximation for the SSW model the exponent $\theta$ and the one-site probability distributions for the bKPZ universality classes can be determined analytically but they differ from the numerical results obtained for the full model.  

As pointed out, extensions of the problem that were already considered are long-range interactions between the substrate and the absorbed particles and the study of the RSOSW model with diffusion of particles. What was not yet studied is nonequilibrium wetting with disorder, which can be very relevant in a experimental situation. Also  relevant in possible experimental realizations, is to consider nonequilibrium wetting in a two-dimensional substrate. The critical exponents of the bKPZ universality class in $d=2$ were not yet determined. Performing Monte Carlo simulations with microscopic models or numerical integration of the Langevin equation at $d=2$ is a trivial task, the main problem is to find some approximative method that can support the results obtained with simulations.

The main challenge, in what we defined here as nonequilibrium wetting, is to observe experimentally the critical behavior obtained theoretically. Any growing interface in the presence of a wall and under nonequilibrium conditions is, in principle, a candidate of an experimental realization of the bKPZ universality classes. Following the discussion in \cite{santos04} good candidates may come from crystal growth and synchronization transitions in extended one-dimensional systems.    

\begin{acknowledgements}

I would like to thank Haye Hinrichsen and M\'{a}rio Jos\'{e} de Oliveira for helpful discussions, collaborations on the present topic and carefully reading the manuscript. Haye Hinrichsen is also acknowledge for fundamental suggestions on writing. The Deutsche Forschungsgemeinschaft is gratefully acknowledge for financial support (HI 744/3-1).

\end{acknowledgements}


\end{document}